\documentstyle[12pt]{article} 

\def\Box{\hbox{$\sqcup$\kern-0.66em\lower0.03ex\hbox{$\sqcap$}}}

\begin{document}
\begin{titlepage}
\begin{flushright}
IFUP--TH 4/96
\end{flushright}
\vskip 1truecm
\begin{center}
\Large\bf
Functional integration on \\
two dimensional Regge geometries\footnote{This work is  supported in part
  by M.U.R.S.T.}.
\end{center}

\vskip 1truecm
\begin{center}
{Pietro Menotti and Pier Paolo Peirano} \\ 
{\small\it Dipartimento di Fisica dell'Universit\`a, Pisa 56100, 
Italy and}\\
{\small\it INFN, Sezione di Pisa}\\
\end{center}
\vskip .8truecm
\begin{center}
January 1996
\end{center}
\end{titlepage} 

\begin{abstract}
By adopting the standard definition of diffeomorphisms for a Regge
surface we give an exact expression of the Liouville action both for
the sphere and the torus topology in the discretized case. The results
are obtained in a general way by choosing the unique self--adjoint
extension of the Lichnerowicz operator satisfying the Riemann--Roch
relation. We also give the explicit form of the integration measure
for the conformal factor. For the sphere topology the theory is
exactly invariant under the $SL(2,C)$ transformations, while for the
torus topology we have exact translational and modular invariance. In
the continuum limit the results flow into the well known expressions.
\end{abstract}

\section{Introduction}

Regge discretized approach to gravity \cite{regge} consists in
replacing regular geometries with piece--wise flat ones with the
curvature confined to $D-2$ dimensional simplices. Apart from
applications to classical gravity such an approach has been considered
as a way to regulate quantum gravity \cite{leehamb}. It has also been
used as the scheme suitable to perform numerical simulations of
quantum gravity (see \cite{pm,catterall} and references therein). In
two dimensions there is the possibility of comparing the results of
Regge gravity to those of the continuum theory.

Most of the discussion on Regge gravity at the quantum level has been
centered on the integration measure \cite{all}, where the most popular
choices have been of the type $\prod_{i} dl_{i} f(l_{i})$, being
$l_{i}$ the bone lengths. On the other hand the continuum approach
\cite{allc}, which was developed from the analogy with gauge theories,
starts from the unique ultra-local, diff--invariant measure, i.e. the
De Witt measure. Given the infinite volume of the diffeomorphism group
a gauge fixing and the evaluation of the related Faddeev--Popov (F.P.)
determinant are required. In particular in $D=2$ this is the only
source of a non trivial dynamics, being the Einstein action a
topological invariant.

In the present paper we shall maintain for the diffeomorphisms the same
meaning as on the continuum. Thus the only difference between the
continuum and the Regge approach will be that in the last case one
restricts the functional integration to the piece--wise flat
surfaces. 

There is a difference between such an approach to gravity and the
usual lattice discretization of gauge theories. In fact in the last
case after discretizing the space--time the action becomes invariant
under a compact group. Thus the imposition of a gauge fixing is not
necessary as one can factorize a finite gauge volume. On the other
hand in order to keep the usual diff-invariance we have to maintain
the description of space--time by the manifold structure
\cite{koba}. Being the symmetry group non compact the gauge fixing
turns out to be necessary.

Given a Regge surface there are many metrics $g_{\mu\nu}$ that
describe such a geometry; the metric has to be given after having
equipped our space--time with a manifold structure, i.e.\ charts with
transition function which are independent of the metric. For example
the metric
\begin{equation}
g^{(1)}_{\mu\nu} = 
\left(    
\begin{array}{cc}
l_{1}^{2} &  \frac{1}{2} (l_{1}^{2} + l_{2}^{2} - l_{3}^{2}) \\
\frac{1}{2} (l_{1}^{2} + l_{2}^{2} - l_{3}^{2}) & l_{2}^{2}
\end{array}
\right) . 
\end{equation}
defined on an open set which includes a triangular simplex with link
lengths $l_1$, $l_2$, $l_3$ and the analogous metric $g^{(2)}_{\mu\nu}$
defined on an open set which covers the adjacent triangular simplex
with link lengths $l_2$, $l_4$, $l_5$ are not compatible on the
intersection region with $l-$independent transition functions.  

As envisaged by Jevicki and Ninomija \cite{jev} we shall maintain the
De Witt metric as the starting point, impose a gauge fixing and
compute the associated integration measure. 
In $D=2$ by far the simplest gauge fixing is the conformal one, as any
metric can be described modulo diffeomorphisms by $g_{\mu\nu} =
\hat{g}_{\mu\nu} e^{2\sigma}$ where $\hat{g}_{\mu\nu}$ is a background
metric depending on the Teichm\"uller parameters.
After imposing the conformal gauge fixing the functional integral
becomes an integral over the conformal factor and on the
Teichm\"uller parameters.  Within our framework one integrates only
over those conformal factors which describe a Regge surface. Thus the
problem reduces to compute the quantities which appear in the continuum
partition function in the case of a Regge geometry, while the
functional integration becomes an integral on a finite number of
parameters which describe such  surfaces. As explained in sect.\
\ref{spheretopology} these parameters will be the positions of the
conical singularities on a coordinate plane, the associated conical
deficits and an overall scale factor.

The first term to be computed is the functional determinant of the
conformal Lichnerowicz operator, i.e.\ the F.P.\ determinant. What is
interesting is that with the above choice of parameters, describing
all the Regge geometries, such a quantity can be obtained exactly in
closed form.  This will be performed by extending the technique
developed by Aurell and Salomonson \cite{aur} for the computation of
the functional determinant of the scalar Laplace--Beltrami operator to
the Lichnerowicz operator that acts on vector fields. Such an
extension in not straightforward \cite{pmppp,proc} because a simple
minded translation of the formulas for the scalar case gives rise to a
wrong result.  The reason is that one has to find out which are the
boundary conditions on the vector field (and on the related traceless
symmetric tensor field) at the singularity suitable for a compact
surface.  In reference \cite{pmppp} the problem of finding the correct
boundary conditions has been solved by regularizing the conical
singularities by means of a smooth geometry and then taking the limit
of vanishing regulator.

Here the problem will be addressed in a completely general way by
looking to all possible self--adjoint extensions of the Lichnerowicz
operator and of the related operator which acts on the traceless
tensor field. The result is that for $\frac{1}{2} < \alpha < 2$
($\alpha$ is the opening of the cone with $\alpha =1$ for the plane)
the imposition of the Riemann--Roch relation for a compact manifold
without boundary, selects a well defined self--adjoint extension which
coincides with the one previously found with the regularization
method. Outside of the interval $(\frac{1}{2}, 2)$ is not possible to
satisfy the Riemann--Roch relation within the realm of $L^{2}$
functions. From the technical viewpoint the calculation of the
determinant is performed similarly as in the continuum, i.e.\ by
taking first a variation of the conformal factor and then integrating
back the result. To this purpose it is necessary to compute the small
time behavior of the heat kernel of the Lichnerowicz operator and of
the associated operator that acts on the traceless symmetric tensors.

We have examined separately the topologies of the sphere and of the
torus. In both cases in the continuum limit the results go over to the
well known expressions.

In the case of the sphere we have explicit invariance under the group
$SL(2,C)$ which corresponds to the six conformal Killing vectors of the
sphere. For the torus we have invariance under translations. 

The expression of the integration measure for the conformal factor
flows directly from the De Witt continuum definition. It is given by
the determinant of a finite dimensional matrix whose elements are
given by integrals which appeared in the old conformal theory
\cite{syma}. One can easily derive the invariance properties
of such a measure. For the sphere topology the measure turns out to be
invariant under $SL(2,C)$ which combined with the invariance of the
action under the same group renders the whole theory $SL(2,C)$ invariant.

The same thing happens for the torus with regard to translations. In
addition here the transformation properties of the action under
modular transformations combined with those of the measure, give rise
to a modular invariant integral of the Liouville action over the
conformal factor, thus assuring the modular invariance of the partition
function.  This procedure provides a non formal proof of the modular
invariance of the theory.

It appears a notable advantage of the geometric nature of the Regge
regulator the fact that such symmetries, like $SL(2,C)$ for the sphere
topology and translation and modular invariance for the torus
topology, are exactly preserved at the discretized level. Obviously
the Regge surface can be equivalently described by the conventional
method of the bone lengths (in fact it is easy to check that one has
the same number of physical degrees of freedom); but the choice we
adopted \cite{foer} appears more suitable for the evaluation of the
functional integral and for the study of its symmetries.

While with such an approach one obtains an action that in the
continuum limit flows in the usual continuum result, it is very hard to
understand how something similar could be obtained using the measure
$\prod_i dl_i f(l_i)$. In fact in $D=2$ the Einstein action is a
topological invariant and thus all the dynamics resides in the
triangular inequalities among the bone lengths. On the other hand for
small variations of the geometry the   Liouville action in the
continuum approach can be approximately computed with one loop
calculation. But at the perturbative level triangular inequalities do
not play any role and thus one does not see how a Liouville action
could emerge.

The paper is structured as follows. In sect.\ \ref{selfaggiunto} we
discuss the self--adjoint extensions of the conformal Lichnerowicz
operator and the related heat kernels; then we impose on them the
restrictions given by the Riemann--Roch relation.
In sect.\ \ref{spheretopology} we apply the above general results to
the sphere topology and give the explicit form of the Liouville action
for a Regge surface. Then we derive the integration
measure of the conformal factor and prove the $SL(2,C)$ invariance of
the theory. In sect.\ \ref{torustopology} we give the Liouville action
for the torus topology and prove the invariance of the functional
integral under modular transformations. In sect.\ \ref{comparison} we
examine briefly the relation of the smooth Liouville action to our
discretized one. In appendix \ref{appc} we give a concise summary of the
continuum gauge fixing procedure, to which we often refer in the
text; in appendix \ref{appa} we give the asymptotic expansion of the
trace of the heat kernels; in appendix \ref{appsphere} we report the
regulator procedure for extracting the boundary conditions at the
singularities and in appendix \ref{appintrep} we write the integral
representation of the heat kernels previously discussed.

\section{Self--adjoint extension of the Lichnerowicz operator}
\label{selfaggiunto}

We need to compute for a Regge manifold
\begin{equation}
\label{siparte}
\displaystyle
  \frac{\det'(P^\dagger P)}{
    \det(\phi_a,\phi_b)\det(\psi_k,\psi_l)}
\end{equation}
(see eq.(\ref{contpart})) where the operator $P$ takes from the 2
dimensional vector field $\xi_{\mu}$ to the traceless symmetric tensor field
$h_{\mu\nu}$
\begin{equation}
  h_{\mu\nu} = \frac{1}{2} \left( \nabla_{\mu}\xi_{\nu} +
  \nabla_{\nu}\xi_{\mu} - \delta_{\mu\nu} \nabla \xi \right)
  = (P\xi)_{\mu\nu} \: .
\end{equation}
Following Alvarez \cite{allc} we go over to the complex formalism
where $\omega= \omega^{1} + i\omega^{2}$, $\xi(\omega, \bar\omega) =
\xi_{\bar\omega} = \xi_{1} + i \xi_{2}$ and $h(\omega, \bar{\omega}) =
h_{\bar{\omega}\bar{\omega}} = h_{11} + i h_{12}$.  The two spaces
$\xi$ and $h$ are equipped with the corresponding invariant metrics, which
in the conformal gauge $g=e^{2\sigma} d\omega \otimes d\bar\omega$
take the form
\begin{equation}
  (\xi^{(1)}, \xi^{(2)}) = \int \! d^{2}\omega \, \bar{\xi}^{(1)} \xi^{(2)}
\end{equation}
and
\begin{equation}
  (h^{(1)}, h^{(2)}) = \int \! d^{2}\omega e^{-2\sigma} \bar{h}^{(1)}
  h^{(2)}.
\label{tenmis}
\end{equation}
It is well known that $P$ acts diagonally on the column vector
$(\xi_{\bar{\omega}} , \xi_{\omega})$ by transforming it into
$(h_{\bar{\omega} \bar{\omega}}, h_{\omega\omega})$

\begin{eqnarray}
\displaystyle
& P 
\left( \begin{array}{c} \xi_{\bar{\omega}} \\ \xi_{\omega}
\end{array} \right) =
\left( \begin{array}{cc}
L & 0 \\ 0 & \bar{L} \end{array} \right)
\left( \begin{array}{c} \xi_{\bar{\omega}} \\ \xi_{\omega}
\end{array} \right) =
\left( \begin{array}{c} h_{\bar{\omega}\bar{\omega}} \\ h_{\omega\omega}
\end{array} \right) & \label{doppiol}
\\ & & \nonumber \\
& P^{\dag}
\left( \begin{array}{c} h_{\bar{\omega}\bar{\omega}} \\ h_{\omega\omega}
\end{array} \right) = 
\left( \begin{array}{cc}
L^{\dag} & 0 \\ 0 & \bar{L^{\dag}} \end{array} \right)
\left( \begin{array}{c} h_{\bar{\omega}\bar{\omega}} \\ h_{\omega\omega}
\end{array} \right) = 
\left( \begin{array}{c} \xi_{\bar{\omega}} \\ \xi_{\omega}
\end{array} \right) \: . & \label{doppiol1}
\end{eqnarray}
In the conformal gauge $L$ and $L^{\dag}$ assume the form
\begin{equation}
  \label{opland}
  L = e^{2\sigma} \frac{\partial}{\partial \bar\omega}
  e^{-2\sigma} \qquad \mbox{and} \qquad 
  L^{\dag} = - e^{-2\sigma} \frac{\partial}{\partial \omega}.
\end{equation}
{}From eqs.(\ref{doppiol}), (\ref{doppiol1}) is clear that $\det'(P^{\dag}P) =
[\det'(L^{\dag}L)]^{2}$ and the determinant of $L^{\dag}L$ is defined
through the $Z$-function technique, i.e.\ $-\log \det'(L^{\dag}L) =
\dot{Z}_{K}(0) \equiv \frac{dZ_{K}(s)}{ds}|_{s=0}$. $Z_{K}(s)$ is
given by the heat kernel of $L^{\dag}L$ as follows
\begin{equation}
  Z_{K}(s) = \frac{1}{\Gamma (s)} \int_{0}^{\infty} \! dt \: t^{s-1} \mbox{Tr
    }'(e^{-tL^{\dag}L})
\end{equation}
where the prime  means exclusion of the zero modes.
The value of $\det' (L^{\dag}L)$ can be written as
\begin{equation}
-\log({\det}'(L^{\dag}L))= \dot{Z}_{K}(0) = \gamma_E Z_{K}(0) + 
{\mbox{Finite}}_{\epsilon \rightarrow 0}
\int^\infty_\epsilon \frac{dt}{t} \mbox{Tr}'(e^{-tL^{\dag}L}) \, .
\end{equation}
The standard procedure is to compute the change of $\dot{Z}_{K}(0)$
under a variation of the conformal factor
\begin{equation}
- \delta \log \left[
  \frac{\det'(L^\dagger L)}{
    \det(\Phi_a,\Phi_b)\det(\Psi_l,\Psi_m)}
\right] 
= \gamma_{E} \delta c_{0}^{K} + 
{\mbox{Finite}}_{\epsilon \rightarrow  0}
\; \mbox{Tr}\:\left[ 4\delta\sigma{\cal K}(\epsilon) 
-2\delta\sigma {\cal H}(\epsilon) \right], 
\label{covva}
\end{equation}
and then integrating back the result.  In the previous equation
$K=L^{\dag}L$, $H=LL^{\dag}$, ${\cal K}$ is the heat kernel of $K$ and
${\cal H}$ is the heat kernel of $H$; $c_{0}^{K}$ is the constant term
in the asymptotic expansion of the trace of the heat kernel ${\cal
K}(t)$ and is related to $Z_{K}(0)$ by
\begin{equation}
c_{0}^{K} = Z_{K}(0) + \mbox{dim}(\mbox{Ker }K). 
\end{equation}
$\Phi_{a}$ and $\Psi_{i}$ are the zero modes of $K$ and $H$
respectively.  The central point in the evaluation of the r.h.s.\ of
eq.(\ref{covva}) will be the knowledge of $c_{0}^{K}$ and of ${\cal
K}(t)$ and ${\cal H}(t)$ on the Regge manifold for small $t$. As is
well known such quantities are local in nature and thus we shall start
by computing them on a single cone.

\subsection{Heat kernels ${\cal K}(\epsilon)$ and ${\cal H}(\epsilon)$
  on a cone}
\hfill \label{sezadj}

In the complex $\omega$ plane a cone is described by the conformal
metric $e^{2\sigma}= c^2 (\omega \bar\omega)^{\alpha -1}$, with
$2\pi\alpha$ the angular aperture and $c$ a normalization constant.
In the polar representation $\omega = r e^{i\phi}$, $L$ and $L^{\dag}$
are given by
\begin{equation}
  \label{opl}
  L = \frac{1}{2} e^{2\sigma} e^{i\phi} \left(
    \frac{\partial}{\partial r} + \frac{i}{r} \frac{\partial}{\partial
      \phi} \right) e^{-2\sigma}
\end{equation}
and
\begin{equation}
  \label{oplc}
  L^{\dag} = - \frac{1}{2} e^{-2\sigma} e^{-i\phi} \left(
    \frac{\partial}{\partial r} - \frac{i}{r} \frac{\partial}{\partial
      \phi} \right).
\end{equation}
By decomposing $\xi$ in angular harmonics $\xi =
\sum_{m=-\infty}^{\infty} e^{im\phi} f_{m}(r)$, the eigenvalue
equation $L^{\dag}L (e^{im\phi} f_{m}(r)) = \lambda^{2} e^{im\phi}
f_{m}(r)$ becomes
\begin{equation}
  \label{conoeq}
  -\frac{c^{-2}}{4} r^{2(1-\alpha)} \left[ \frac{d^{2}}{dr^{2}} + (3 -
    2\alpha) \frac{1}{r} \frac{d}{dr} -\frac{1}{r^{2}} (m^{2}
    +2(\alpha -1)m) \right] f_{m} = \lambda^{2} f_{m}
\end{equation}
which is solved by $f_{m} = r^{\alpha -1} J_{\pm \nu}
(\frac{2c\lambda}{\alpha}r^{\alpha})$ with $\nu= \frac{m+\alpha-1}{\alpha}$, or
linear combination thereof.

The condition of $L^{2}(r\,dr)$ integrability at the origin dictates
the choice 
\begin{equation}
f_{m} = \left\{ \begin{array}{ll} \displaystyle
r^{\alpha -1} J_{\nu} (\frac{2c\lambda}{\alpha}r^{\alpha}) & \qquad
\mbox{for } \nu > -1 \\ \quad & \quad \\ \displaystyle
r^{\alpha -1} J_{-\nu} (\frac{2c\lambda}{\alpha}r^{\alpha}) & \qquad
\mbox{for } \nu < 1
\end{array} \right. \quad .
\end{equation}
If only one of the two inequalities is satisfied there is no
ambiguity, else we have that any linear combination of $J_{\nu}$ and
$J_{-\nu}$ is $L^{2}$-integrable.  This gives rise to the problem of
the choice of the domain of self--adjointness of the operator
$L^{\dag}L$. Moreover as $L^{\dag}L$ originates from $(L\xi, L\xi)$ we
shall require $L^{\dag}L$ to be really the product of an operator $L$
and of its adjoint $L^{\dag}$, with the ensuing restrictions on the
domains $D(L)$ and $D(L^{\dag})$.

We shall start by looking at the domain of self--adjointness of
$(L^{\dag}L)_{m}$ ($L^{\dag}L$ restricted to the partial wave $m$)
with $-1 < \frac{m +\alpha -1}{\alpha} < 1$. First we define
$K = (L^{\dag}L)_{m}$ as a closed symmetric operator. 
The domain of $K$, $D(K)$ will be defined by the functions $\xi\in
L^2(r~dr)$ with two derivatives and such that
\begin{equation}
\lim_{r\rightarrow 0}~ r^{-2\alpha+1+\epsilon} \xi =0 \, , \qquad \quad
\lim_{r\rightarrow 0}~ r^{-2\alpha+2+\epsilon} \xi' =0 
\label{comporig}
\end{equation}
for any $\epsilon>0$.

We want to find $D(K^{\dag})$. We have
\begin{eqnarray}
\displaystyle
  \lefteqn{ \int \! r\, dr \; \bar{\eta} (L^{\dag}L)_{m}\xi }  & & \\ &
  & \! \! \! \! \! \! \! \!
  = -\frac{1}{4} \int \! rdr \, [e^{-2\sigma} (\frac{\partial}{\partial
    r} + \frac{m + 1}{r}) e^{2\sigma}(\frac{\partial}{\partial r} -
  \frac{m}{r}) e^{-2\sigma} \bar{\eta}] \xi   
  + \lim_{r\rightarrow 0} [r ( 
\bar{\eta} \frac{\partial}{\partial r} (e^{-2\sigma}\xi)  - 
\frac{\partial}{\partial r}(e^{-2\sigma} \bar{\eta})\xi)].  
\nonumber
\end{eqnarray}
$D(K^{\dag})$ is given by the set of the $L^2(r~dr)$ functions $\eta$  with two
derivatives, and for which
\begin{equation}
  \lim_{r\rightarrow 0}~ r e^{-2\sigma} (\bar{\eta} \xi' -
  \bar{\eta'} \xi) =0 \, , \qquad \forall \xi \in D(K).
\end{equation}
{}From the definition of $D(K)$ it follows that $D(K^{\dag}) \supset
D(K)$. In particular $D(K^{\dag})$ includes the $L^2(r dr)$ functions
with two derivatives which in a neighborhood of the origin are equal
to $r^{-1 + \epsilon}$ with $\epsilon>0$ and this implies according
to eq.(\ref{comporig}) that $D((K^{\dag})^{\dag}) = D(K)$, in other
words $K$ defined on $D(K)$ is a closed operator.

$D(K^{\dag})$ properly contains $D(K)$ and thus $K$ is not self--adjoint and
now we look for all possible self--adjoint extensions of it. As $K$
defined on $D(K)$ is a closed operator we can apply the standard theory
of self--adjoint extensions.
We must
look for the $L^{2}(r\,dr)$ solutions of $(K^{\dag} \pm i) \xi =0$. As
$K$ is a real operator the dimension of Ker$(K^{\dag} - i)$ equals
the dimension of Ker$(K^{\dag} + i)$, which assures that there exist 
self--adjoint extensions of $K$.

We have only one $L^{2}(r\,dr)$ solution of $(K^{\dag} - i) \xi_{0}=0$ and it
is given by
\begin{equation}
  \xi_{0} = r^{\alpha -1} \left( J_{-\nu} (\frac{2c}{\alpha}
    e^{i\frac{\pi}{4}} r^{\alpha}) - e^{-i\nu\pi} J_{\nu}
    (\frac{2c}{\alpha} e^{i\frac{\pi}{4}} r^{\alpha}) \right) =
i \sin(\nu\pi)  
r^{\alpha -1} H^{(1)} (\frac{2c}{\alpha} e^{i\frac{\pi}{4}}
  r^{\alpha})
\end{equation}
while the solution of $(K^{\dag} + i) \xi=0$ is obviously given by its
complex conjugate.

The general theory of the self--adjoint extensions of a symmetric
operator \cite{ds} tell us that we have as many extensions as the
unitary maps between Ker$(K^{\dag} -i)$ and Ker$(K^{\dag}
+i)$. Therefore in principle, if for a given $\alpha$ there are $l$
values of $m$ for which $-1<\nu<1$, then we have an $l^{2}$--dimensional
family of self--adjoint extensions.

On the other hand if we want to preserve invariance under rotations
around the tip of the cone, we can mix only solution of Ker$(K^{\dag}
-i)$ and Ker$(K^{\dag} +i)$ with the same angular momentum. Thus we
are left only with an $l$--dimensional family of self--adjoint extensions.

The domain of such self--adjoint extension is given for each partial
wave by $D(K) \oplus (e^{i\theta} \xi_{0} + e^{-i\theta} \bar{\xi}_{0})$.
This is completely equivalent to add to the initial domain $D(K)$ the
$L^{2}(r\,dr)$ functions which at the origin behave like $r^{\alpha -1}
(a r^{-\alpha \nu} + b r^{\alpha\nu})$ with $\frac{a}{b}$ real and
fixed.  One can easily check that despite having introduced a
singular behavior at the origin, in passing form $(\eta, K\xi)$ to
$(K\eta, \xi)$ the integrations by parts are carried through
without leaving boundary terms.

We now impose that $K=L^{\dag} \cdot L$, i.e.
\begin{equation}
  \label{prodreq}
  \mbox{Im}(L) \subseteq \mbox{D}(L^{\dag}).
\end{equation}
We shall show that eigenfunctions of the type $a J_{\nu} + b J_{-\nu}$
violate this requirement. In fact given
\begin{equation}
  \xi_{\lambda} = a \left( \frac{\alpha}{2\lambda c} \right)^{\nu}
  r^{\alpha -1} J_{\nu} \left( \frac{2\lambda c}{\alpha} r^{\alpha}
  \right) + b \left( \frac{\alpha}{2\lambda c} \right)^{-\nu} r^{\alpha
    -1} J_{-\nu} \left( \frac{2\lambda c}{\alpha} r^{\alpha} \right)
\end{equation}
we have
\begin{equation}
  (\xi_{\lambda'}, K \xi_{\lambda}) =
  \int_{0}^{\infty} \! dr\,r\; \overline{(L\xi_{\lambda'})} e^{-2\sigma}
  (L\xi_{\lambda}) + \frac{1}{2} ab(1-\alpha -m).
\label{partii}
\end{equation}
The requirement $L\xi \in L^{2} (e^{-2\sigma}r\,dr)$, i.e. that the
integral on the r.h.s.\ of eq.(\ref{partii}) converges, imposes that for
$0< \nu < 1$, $b$ must be taken equal to zero. Thus the values of $\nu$
for which the choice is ambiguous is $(-1,0)$. Furthermore from
eq.(\ref{partii}) the imposition that $(\xi_{\lambda'}, K\xi_{\lambda}) =
(L\xi_{\lambda'}, L\xi_{\lambda})$ imposes that either $a$ or $b$
equals zero.

In conclusion the requirement of self--adjointness of $K$, imposes on
each partial wave for which two independent $L^{2}$ eigenfunction exist, a
universal linear combination of them, while the requirement that
$K=L^{\dag}\cdot L$ reduces the choice either to the regular or to the
irregular solution. Such a choice exists only for $-1 < \nu < 0$.

We now examine how the imposition of the Riemann--Roch theorem is able
to further restrict the choice of the self--adjoint extension of
$K$. The Riemann--Roch theorem for a 2--dimensional compact surface
states that
\begin{equation}
  \label{riroch}
  \mbox{dim }(\mbox{ker } P) - \mbox{dim }(\mbox{ker }P^{\dag}) = 3
  \chi
\end{equation}
being $\chi=2 -2h$ the Euler characteristic of the surface of genus
$h$.  With $L$ and $L^{\dag}$ referred to the whole manifold we have
\begin{equation}
  c_{0}^{K} =  Z_{K}(0) + \mbox{dim }(\mbox{ker } L) \qquad c_{0}^{H} = 
  Z_{H}(0) + \mbox{dim }(\mbox{ker } L^{\dag}).
\end{equation}
We recall that $K=L^{\dag}L$, $H=LL^{\dag}$ and $c_{0}^{K}$,
$c_{0}^{H}$ are the constant coefficients in the asymptotic expansions
of the trace of the heat kernels ${\cal K}(t) = e^{-t L^{\dag}L}$ and 
${\cal H}(t) = e^{-t LL^{\dag}}$ for small $t$.

The spectra of $K$ and $H$ coincide except for the zero modes, hence
$Z_{K}(0) = Z_{H}(0)$. Since the same treatment can be applied to the
operators $\bar{L}$ and $\overline{L^{\dagger}}$ acting on $\xi_\omega$ and
on $h_{\omega\omega}$ with the same results, we have
\begin{equation}
  2( c_{0}^{K} - c_{0}^{H} ) = \mbox{dim }(\mbox{ker }(P^\dagger P)) -
\mbox{dim }(\mbox{ker }(PP^\dagger)) = 3\chi.
\end{equation}
So we must check whether on our singular manifold the relation $2(
c_{0}^{K} - c_{0}^{H} ) = 3\chi$ is satisfied. Due to the local nature
of the coefficients in the asymptotic expansion of the heat kernel, in
order to respect the Riemann-Roch result, we need for a single conical
singularity 
\begin{equation}
  \label{singsing}
  2(c^K_{0i} - c^H_{0i} ) = 3 (1 -\alpha_{i}).
\end{equation}
We shall see that this requirement selects unambiguously the domain of
self-adjointness of $K$ for $\frac{1}{2} < \alpha < 2$.

We note that the choice $\nu_{m} = \frac{m + \alpha -1}{\alpha}$ for
$m \geq 0$, $\nu_{m} = - \frac{m + \alpha -1}{\alpha}$ for $m < 0$, in
the interval $\frac{1}{2} < \alpha < 2$ satisfies the discussed
requirement of self-adjointness of $K = L^{\dag} \cdot L$, as it
chooses functions which are $L^{2}$ and are never mixtures of a
regular and a singular solution.

In appendix \ref{appa} we report the calculation of $c_{0}^{K}$ for a
generic phase shift $\delta$ (i.e.\ $\nu_{m} = \frac{m +
\delta}{\alpha}$, $m \geq 0$; $\nu_{m} = - \frac{m -\delta}{\alpha}$,
$m<0$). With the above choice $\delta= \alpha -1$, we obtain
\begin{equation}
  \label{ckris}
  c_{0}^{K} = \frac{1 - \alpha^{2}}{12\alpha} + \frac{(\alpha -
    1)(\alpha - 2 )}{2\alpha}.
\end{equation}
Using
\begin{equation}
  \label{rules}
  L \left( e^{im\phi} r^{\alpha -1} J_{\nu_{m}} ( \frac{2\lambda
      c}{\alpha} r^{\alpha} ) \right) = \mbox{const} \, e^{i(m+1)\phi}
  r^{2(\alpha -1)} J_{\gamma_{m}} ( \frac{2\lambda
    c}{\alpha}r^{\alpha} )
\end{equation}
with $\gamma_{m} = \nu_{m} + 1$ for $\nu_{m} = \frac{m + \alpha -
  1}{\alpha}$ and $\gamma_{m} = \nu_{m} - 1$ for $\nu_{m} = - \frac{m
  + \alpha -1}{\alpha}$. $c_{0}^{H}$ is given by
eq.(\ref{c0h}) i.e.
\begin{equation}
  \label{chris}
  c_{0}^{H} = \frac{1 - \alpha^{2}}{12\alpha} + \frac{(2\alpha -
    1)(2\alpha - 2 )}{2\alpha}.
\end{equation}
Taking the difference we have
\begin{equation}
  \label{rragr}
  2 (c_{0}^{K} - c_{0}^{H}) = 3 (1 - \alpha)
\end{equation}
in agreement with the Riemann--Roch theorem.  The other possible
self--adjoint extensions of $K = L^{\dag} \cdot L$ differ from the
previously described one by replacing for $-1 < \nu < 0$ a singular
(regular) solution with a regular (singular) one.

 For $\frac{1}{2} < \alpha < 1$ we saw that the choice of the singular
eigenfunction satisfies the
Riemann-Roch relation. Replacing it with
the regular one amounts to change a term in
the sum eq.(\ref{sommappa}), i.e.\ the term
corresponding to $m=0$ given by
\begin{equation}
  -\frac{1}{2} \left[ \left( \frac{\alpha - 1}{\alpha} \right)^{1 -2s}
    + s \left( \frac{\alpha - 1}{\alpha} \right)^{-1 -2s}B_{2}
  \right]
\end{equation}
with
\begin{equation}
  -\frac{1}{2} \left[ \left( \frac{1 - \alpha}{\alpha} \right)^{1 -2s}
    + s \left( \frac{1 - \alpha}{\alpha} \right)^{-1 -2s}B_{2}
  \right]
\end{equation}
which in the limit $s \rightarrow 0$ gives
\begin{equation}
  \label{deltack}
  \Delta c_{0}^{K} = {c'}_{0}^{K}- c_{0}^{K} = \frac{ \alpha - 1}{\alpha}.
\end{equation}
The change in the eigenfunctions of $K$ determines a well defined
change in the eigenfunctions of $H$ according to the eq.(\ref{rules}),
giving
\begin{equation}
  \label{deltach}
  \Delta c_{0}^{H} = {c'}_{0}^{H}-  c_{0}^{H} = \frac{\alpha -
  1}{\alpha} + 1 \, .
\end{equation}
Taking the difference we have
\begin{equation}
 \label{violar} 
 \Delta (c_{0}^{K} - c_{0}^{H}) = -1
\end{equation}
thus giving rise to a violation of the Riemann-Roch relation. Similarly
for $1 < \alpha < \frac{3}{2}$ we find that the alternative choice
gives
\begin{equation}
  \label{deltac}
  \Delta (c_{0}^{K} - c_{0}^{H}) = 1.
\end{equation}
For $\frac{3}{2} < \alpha < 2$, we have 3 alternative possibilities
due to the fact that we have two angular momenta $m=-1$ and $m=-2$
with two acceptable eigenfunctions. Substituting in our expression one
of the two eigenfunctions with the alternative one, we obtain the same
violation of eq.(\ref{deltac}), while substituting both, the result is
\begin{equation}
  \Delta (c_{0}^{K} - c_{0}^{H}) = 2.
\end{equation}
In conclusion in the interval $\frac{1}{2} < \alpha < 2$ the
imposition of the Riemann--Roch relation singles out a unique
self--adjoint extension of $K$ and $H$. We come now to discuss $\alpha
< \frac{1}{2}$ and $\alpha > 2$. For $\alpha < \frac{1}{2}$ and $m=0$
we have a unique $L^{2}$ eigenfunction which corresponds to choosing
$J_{-\nu_0}$, which as we have just seen, see eq.(\ref{violar}), violates the
Riemann--Roch relation.

For $\alpha > 2$ the requirement of $L^{2}$ summability on the
eigenfunctions requires some terms with index $\frac{n -
\delta}{\alpha}$ in the second sum appearing in eq.(\ref{sommappa}),
to be substituted by the corresponding ones with index $\frac{\delta -
n}{\alpha}$. Each of these shifts gives rise to a violation of $1$ in
the Riemann--Roch relation. In addition for the angular momenta for
which $1 - 2 \alpha < m < 1 - \alpha$, where two independent $L^{2}$
function exist, each further allowed shift from the second to the
first sum gives rise to an additional violation of $1$.  Thus outside
the interval $\frac{1}{2} < \alpha < 2$ is not possible to satisfy the
Riemann--Roch relation within the realm of $L^2$-functions. In
appendix \ref{appsphere} we shortly report the calculation with the
regulator technique, which gives the same result.

\section{Sphere topology}
\label{spheretopology}

We recall that in two dimensions, modulo diffeomorphisms, every metric
can be given in terms of a background metric of constant curvature
multiplied by a conformal factor $g_{\mu\nu}= e^{2\sigma}\hat
g_{\mu\nu}$. We start with the topology of the sphere. The usual
choice is to describe it through a stereographic projection on the
plane, with $\hat{g}_{\mu\nu} = \delta_{\mu\nu}$. Then for a Regge
geometry we have
\begin{equation}
e^{2\sigma} = e^{2\lambda_{0}} \prod_{i=1}^{N} | \omega - \omega_{i}
|^{2(\alpha_{i} -1)}, \quad \sigma \equiv \sigma(\omega ; \lambda_{0},
\omega_{i}, \alpha_{i} )= \lambda_0+\sum_i(\alpha_i-1)\log|\omega-\omega_i|
\end{equation}
with the restriction $\sum_{i=1}^{N} (1 - \alpha_{i}) = 2$, i.e.
the sum of the deficits must be equal to the Euler characteristic.
Due to the presence of 6 conformal Killing vectors for a manifold with
the topology of a sphere, the gauge fixing is not complete, i.e.\ two
different $\sigma$ related by an $SL(2,C)$ transformation, describe
the same geometry. The transformation is
\begin{equation}
  \omega' = \frac{\omega a + b}{\omega c + d} \; , \quad  \omega =
  \frac{\omega'd - b}{- \omega' c + a} \; , \quad  ad - bc = 1 
  \label{sl1}
\end{equation}
and $\sigma$ goes over to 
\begin{eqnarray} 
\displaystyle
\lefteqn{\sigma'(\omega';\lambda_{0}, \omega_{i}, \alpha_{i}) \equiv
\sigma(\omega(\omega'); \lambda_{0}, \omega_{i}, \alpha_{i}) + \log |
\frac{\partial \omega}{\partial \omega'} | } & & \\
& & = \lambda_{0} + \sum_{i} (\alpha_{i} - 1 ) \log | \frac{\omega' d - b}{
  - \omega' c + a}
-\omega_{i}| -2 \log |c \omega' -a| \nonumber \\
& & = \lambda_{0} + \sum_{i=1}^{N} (\alpha_{i} - 1)  \log \frac{|
  \omega' d - b 
- \omega_{i} (a -  \omega' c) |}{|\omega_{i} c + d |} +
\sum_{i=1}^{N} (\alpha_{i} -1 )\log |\omega_{i} c + d|  \nonumber \\
& & = \lambda_{0} + \sum_{i=1}^{N} (\alpha_{i} - 1) \log |\omega' -
\frac{ \omega_{i} a  + b}{ \omega_{i} c + d}| + 
\sum_{i=1}^{N} (\alpha_{i} -1 )\log |\omega_{i} c + d| \nonumber
\end{eqnarray}
having used $\sum_{i=1}^{N} (1 - \alpha_{i}) = 2$.
The new conformal factor is given by 
\begin{equation}
\sigma'(\omega'; \lambda_{0}, \omega_{i}, \alpha_{i}) =
\sigma(\omega';  \lambda_{0}', \omega_{i}', \alpha_{i}')
\end{equation}
with
\begin{equation}
\label{conftras}
\lambda_{0}' = \lambda_{0} + \sum_{i=1}^{N} (\alpha_{i} -1 ) \log
|\omega_{i} c + d|, \quad
\displaystyle \omega_{i}' = \frac{a\omega_{i} + b}{c\omega_{i} + d },
\qquad \alpha_{i}' = \alpha_{i} \, .
\end{equation}
Under such transformation the area $A$
\begin{equation}
A=e^{2\lambda_0}\int d^2\omega |\omega-\omega_i|^{2(\alpha_i-1)}.
\end{equation}
being a geometric invariant is left unchanged. We notice that the
number of physical degrees of freedom is the same as the number of
links in the usual parameterization of a Regge surface. In fact from
the Euler relation $F+V = H+2$ with $H=\frac{3}{2} F$ we get $H=3V
-6$, where $-6$ corresponds to the $6$ conformal Killing vectors of
the sphere.

In the neighborhood of $\omega_{i}$ the conformal factor can be
rewritten as  
\begin{equation}
|\omega - \omega_i|^{2(\alpha_i-1)} e^{2\lambda_{i}}\quad
\mbox{with}\quad
\lambda_i = \lambda_0 + \sum_{j\neq i} (\alpha_j-1)\log |\omega_{j}
- \omega_{i}|. 
\end{equation}
In working out the r.h.s. of eq.(\ref{covva}) it is simpler to use cartesian
coordinates \cite{aur} given by
$z=\frac{e^{\lambda_{i}}}{\alpha_{i}} (\omega
-\omega_{i})^{\alpha_{i}}$ with $\frac{dz}{d\omega} = e^{\lambda_{i}}
(\omega - \omega_{i})^{\alpha_{i}-1}$. To a variation $\delta
\lambda_{i}$ and $\delta \alpha_{i}$ there corresponds a variation in
$\sigma$ 
\begin{equation}
\delta \sigma(z,\bar z) =\log|{dz'\over dz}|= (\delta \lambda_{i} - 
\lambda_{i} 
\frac{\delta \alpha_{i}}{ \alpha_{i}})
+ \frac{\delta \alpha_{i}}{\alpha_{i}} \log(\alpha_{i}|z|)
\end{equation}
and substituting in eq.(\ref{covva}) we have, taking into account that
on the sphere there are no Teichm\"uller parameters
\begin{eqnarray}
    \label{beldelta}
    \lefteqn{
    - \delta \log \frac{{\det}'(L^{\dag}L)}{\det(\Phi_{a}, \Phi_{b})}
    = \gamma_{E} \delta c_{0}^{K} + \sum_{i} \left\{ (\delta
    \lambda_{i} -\lambda_{i} \frac{\delta \alpha_{i}}{\alpha_{i}}) [ 4
    c_{0\ i}^{K} - 2 c_{0\ i}^{H}]  \right.} & &   \\ & & 
    \! \! \! \! \! \! \! \displaystyle \left. +
    {\mbox{Finite}}_{\epsilon \rightarrow 0} \left[ 4
    {\displaystyle\frac{\delta \alpha_{i}}{\alpha_{i}}} \int\!d^{2}x\,
    \log (\alpha_{i}|{\bf x}|) \, {\cal
    K}_{\alpha_{i}}({\bf x}, {bf x},\epsilon) -2
    {\displaystyle\frac{\delta \alpha_{i}}{\alpha_{i}}} \int\!d^{2}x\,
    \log (\alpha_{i}|{\bf x}|) \, {\cal H}_{\alpha_{i}}({\bf x},
    {\bf x},\epsilon) \right] \right\}, \nonumber
\end{eqnarray}
where from eqs.(\ref{ckris}), (\ref{chris})
\begin{equation}
\label{beldiff}
4 c_{0\ i}^{K} - 2 c_{0\ i}^{H}={13\over 6}({1\over \alpha_i}-\alpha_i).
\end{equation}
A differential of this structure can be integrated. In fact the
expression in the square brackets depends only on the $\alpha_i$ separately
while one can write \cite{aur}
\begin{equation}
\label{belwrite}
(\delta \lambda_{i} - \lambda_{i} \frac{\delta
\alpha_{i}}{\alpha_{i}})({1\over \alpha_i}-\alpha_i)
= \delta \left[ \left( \frac{1}{\alpha_{i}} -\alpha_{i} \right)
  \lambda_{i} \right] + 2 \delta \alpha_{i} \lambda_{i}  
\end{equation}
and as $\sum_{i=1}^{N} \delta\alpha_{i} = 0$
\begin{eqnarray}
\sum_{i} \delta\alpha_{i} \lambda_{i} &=& \sum_{i} \delta \alpha_{i}
\sum_{j \neq i} (\alpha_{j} -1) \log |\omega_{i} - \omega_{j}|    \\
&+& \sum_{j} (\alpha_{j} -1) \sum_{i \neq j} (\alpha_{i} -1) \delta (\log
|\omega_{i} - \omega_{j}|) \nonumber
\end{eqnarray}
due to the antisymmetry of $\delta(\log |\omega_{i} - \omega_{j}|)$ in
$i,j$. Thus
\begin{equation}
\sum_{i=1}^{N} \delta \alpha_{i} \lambda_{i} = \sum_{i=1}^{N}
(\alpha_i-1)( \delta \lambda_{i} - \delta \lambda_{0})
\end{equation}
and then
\begin{equation}
2 \sum_{i=1}^{N} \delta \alpha_{i} \lambda_{i} = \sum_{i=1}^{N} \delta
\alpha_{i} \lambda_{i} + \sum_{i=1}^{N} (\alpha_{i} - 1) ( \delta
\lambda_{i} - \delta \lambda_{0}) =
\delta \left[ \sum_{i=1}^{N} (\alpha_{i} -1)(\lambda_{i} -\lambda_{0})
\right].
\end{equation}
All the above reasonings refer to the operator $L^{\dag} L$ acting on the
field $\xi$. The same treatment is to be applied to the field
$\xi_\omega$ and so one has to multiply the result by a factor
$2$. The final result for the determinant is 
\begin{eqnarray}
  \label{primoris}
  \lefteqn{
  \log \sqrt{\frac{{\det}'(P^{\dag}P)}{\det(\phi_{a}, \phi_{b})}}} &
  & \\ & & 
  \displaystyle =\frac{26}{12} \left\{ \sum_{i,j\neq i}
    \frac{(1-\alpha_{i})(1-\alpha_{j})}{\alpha_{i}} \log |w_{i} -
    w_{j}| + \lambda_{0} \sum_{i} (\alpha_{i} - \frac{1}{\alpha_{i}})
    - \sum_{i} F(\alpha_{i}) \right\} \nonumber
\end{eqnarray}
where $F(\alpha)$ is given by $\gamma_E c^K_0(\alpha)$ added to
the primitive of 
\begin{equation}
\label{daprima}
{\mbox{Finite}}_{\epsilon \rightarrow 0}
 \left[
\frac{4}{\alpha} \int \! d^{2}x \, \log(\alpha | {\bf x}|) {\cal
  K}_{\alpha}({\bf x}, {\bf x};
\epsilon) - \frac{2}{\alpha} \int \! d^{2}x \, \log(\alpha | {\bf x}|)
{\cal H}_{\alpha}({\bf x}, {\bf x}; \epsilon) \right]
\end{equation}
(see appendix \ref{appintrep} for an integral representation of these
terms). By direct substitution one verifies that eq.(\ref{primoris})
is invariant under the $SL(2,C)$ transformations (\ref{conftras}).

We notice that apart from the term $\sum_{i=1}^{N} F(\alpha_{i})$
eq.(\ref{primoris}) is exactly $-26$ times the conformal anomaly for the
scalar field as computed by Aurell and Salomonson \cite{aur}.  In the
continuum limit $N \rightarrow \infty$, the $\omega_{i}$ become dense
and the $\alpha_{i} \rightarrow 1$, always with $\sum_{i=1}^{N}
(1-\alpha_{i})=2$. In such a limit $\sum_{1=1}^{N} F(\alpha_{i})$ goes
over to the topological invariant $N\, F(1) - \chi F'(1)$, while the
remainder goes over to the well known continuum expression. In fact
we have
\begin{equation}
\frac{1}{2\pi} \log | \omega - \omega' | = \frac{1}{\Box} (\omega, \omega')
\end{equation}
and for any region $V$ of the plane $\omega$
\begin{equation}
\int_{V} \! d^{2}\omega \, e^{2\sigma} R = -2 \int_{V} \! d^{2}\omega
\, \Box \sigma = 4\pi \sum_{i: \omega_{i} \in V} (1 -\alpha_{i}).
\end{equation}
Thus the r.h.s. of eq.(\ref{primoris}) goes over to 
\begin{equation}  
\displaystyle
\label{contliouv}
\frac{26}{96\pi} \left\{\int  d^{2}\omega \, d^{2}\omega' \;(\sqrt{g}
  R)_{\omega }
\frac{1}{\Box}(\omega, \omega')(\sqrt{g} R)_{\omega'}   
 - 2 (\log {A\over A_0}) 
\int \: d^{2} \omega\,\sqrt{g} R \right\}
\end{equation}
where $A_0$ is the value of the area for $\lambda_0=0$.

\subsection{Integration measure for the conformal factor}
\hfill

We work out the functional integration measure ${\cal D}[\sigma]$
appearing in appendix \ref{appc} in the Regge framework.  The
distance between two nearby configurations $\sigma$ and $\sigma +
\delta \sigma$ is given by
\begin{equation}
  (\delta \sigma, \delta \sigma) = \int \! d^{2}\omega \, e^{2\sigma} \,
  \delta \sigma \, \delta \sigma \, .
\label{confmetr}
\end{equation}
Such an expression is a direct outcome of the original De-Witt
measure (\ref{dewittm}).

{}From eq.(\ref{confmetr}) it follows that having parameterized the
Regge surface by means of the $3N$ variables $p_{i}$
\begin{equation}
  \{ p_{1},\ldots, p_{3N} \} \equiv \{ \omega_{1,x}, \omega_{1,y},
  \omega_{2,x}, \omega_{2,y}, \ldots, \omega_{N,x}, \omega_{N,y},
  \lambda_{0}, \alpha_{1}, \alpha_{2}, \ldots, \alpha_{N-1} \}
  \nonumber
\end{equation}
${\cal D}[\sigma]$ is given by
\begin{equation}
  {\cal D}[\sigma] = \prod_{k=1}^{N} d^{2}\omega_{k} \;
  \prod_{i=1}^{N-1} d\alpha_{i} d\lambda_{0} \, \sqrt{\det J}
\end{equation}
being $J$ the $3N \times 3N$ matrix
\begin{equation}
  \label{jac}
  J_{ij} = \int d^{2}\omega \, e^{2\sigma} \, \frac{\partial
    \sigma}{\partial p_{i} } \frac{\partial \sigma}{\partial p_{j}},
\end{equation}
with $\alpha_{N} = \sum_{i=1}^{N-1} (1 -\alpha_{i}) -1$.
We recall the expression for $\sigma$
\begin{equation}
  \label{conffact}
  e^{2\sigma} = e^{2\lambda_{0}} \prod_{i=1}^{N} | \omega
  -\omega_{i}|^{2(\alpha_{i} -1)}.
\end{equation}
Eq.(\ref{jac}) can be given a more transparent form by doubling the
number of variables, i.e.\ using $\omega_{k}, \tilde{\omega}_{k},
\lambda_{0}, \tilde{\lambda}_{0}, \alpha_{i}, \tilde{\alpha}_{i}$ and
introducing a new conformal factor
\begin{equation}
  \label{newfact}
  e^{\sigma(p) + \sigma (\tilde{p})} = e^{\lambda_{0} + \tilde{\lambda}_{0} }
  \prod_{i=1}^{N} | \omega -\omega_{i}|^{(\alpha_{i} -1)} | \omega
  - \tilde{\omega}_{i}|^{(\tilde{\alpha}_{i} -1)}
\end{equation}
and computing
\begin{equation}
  \label{darea}
  \tilde{A} = \int d^{2}\omega e^{\sigma (p) + \sigma(\tilde{p})}
\end{equation}
which is the area of the Regge manifold described by the $6N$
parameters $p$ and $\tilde{p}$.

It is easily verified that
\begin{equation}
  \label{jac2}
  J_{ij} = \left[ \frac{\partial^{2} \tilde{A}}{\partial p_{i} \partial
      \tilde{p}_{j}} \right]_{p=\tilde{p}}.
\end{equation}
Being $\sigma = \lambda_{0} + \frac{1}{2} \sum_{i=1}^{N} (\alpha_{i}
-1) \log |\omega - \omega_{i} |^{2}$ we obtain
\begin{eqnarray}
  \label{var}
  \frac{\partial \sigma}{\partial \omega_{i,x}} = (\alpha_{i} - 1 )
  \frac{\omega_{i,x} - \omega_{x}}{|\omega_{i,x} - \omega_{x}|^{2}} &
  \nonumber
  \\ \frac{\partial \sigma}{\partial \alpha_{i}} = \frac{1}{2} \log |
  \omega_{i} - \omega |^{2} - \frac{1}{2} \log | \omega_{N} - \omega
  |^{2}, & i \leq N - 1 \\ \frac{\partial \sigma}{\partial
  \lambda_{0}} = 1 \, , & \nonumber 
\end{eqnarray}
that substituted in eq.(\ref{jac}) with $e^{2\sigma}$ given by
eq.(\ref{conffact}) give all elements $J_{ij}$.

Each row $J_{\omega_{i,x}, p_{j}}$ contains a factor $\alpha_{i} - 1$.
Due to the multi--linear property of the determinant in the rows,
$\det J$ will factorize a factor $\prod_{i=1}^{N}(\alpha_{i} - 1 )^{2}$
i.e.\ 
\begin{equation}
  \label{detvan}
  \det J = \prod_{i=1}^{N}(\alpha_{i} - 1)^{2} F(p) \, .
\end{equation}

The vanishing of $\det J$ whenever an $\alpha_{i}$ equals $1$ is
expected from the fact that in such situation the position of the
vertex $i$ is irrelevant in determining the metric.

A measure of the structure
\begin{equation}
  \label{meas}
  \prod_{i=1}^{N} d\omega_{i,x} d\omega_{i,y} \, \prod_{j=1}^{N-1}
  d\alpha_{i} \, d\lambda_{0} \, \sqrt{\det (J)}
\end{equation}
with $J$ given by eq.(\ref{jac2}) is invariant under the $SL(2,C)$
transformations
\begin{eqnarray}
  \omega_{i}' = \frac{\omega_{i} a + b}{\omega_{i} c + d} & \nonumber \\ 
  \label{misconfinv}
 \lambda_{0}' = \lambda_{0} + \sum_{i=1}^{N} (\alpha_{i} - 1)
  \log | \omega_{i} c + d | & \\ \alpha_{i}' = \alpha_{i}, & \qquad ad -
  bc = 1. \nonumber
\end{eqnarray}
In fact $A$ is an invariant
\begin{equation}
  A(p) = A(p'), \mbox{\ \ \ \ \ \ } \tilde{A}(p,\tilde{p}) =
  \tilde{A}(p',\tilde{p}').
\end{equation}
Then
\begin{equation}
  \frac{\partial^{2} \tilde{A}}{\partial p_{i} \partial \tilde{p}_{j}}
  \, dp_{i} \, 
  d\tilde{p}_{j} = \frac{\partial^{2} \tilde{A}}{\partial p_{k}' \partial
    \tilde{p}_{l}'} \left( \frac{\partial p_{k}'}{\partial p_{i}}
  \right) \left( 
  \frac{\partial \tilde{p}_{l}'}{\partial \tilde{p}_{j}} \right) \,
  dp_{i} \, d\tilde{p}_{j} 
\end{equation}
and for $p=\tilde{p}$ we obtain
\begin{equation}
\label{dettrans}
  \sqrt{\det J} = \sqrt{\det J'} \; \det \left(
    \frac{\partial p_{k}'}{\partial p_{i}} \right)
\end{equation}
which proves that $\sqrt{\det J} \; \prod_{i=1}^{3N} dp_{i} =
\sqrt{\det J'} \; \prod_{i=1}^{3N} dp_{i}'$.

We notice that all $J_{ij}$ are given by convergent integrals except
those involving two $\omega_{i}$ with the same indexes, which
converge only for $\alpha_{i} > 1 $. For example we
have
\begin{equation}
  \label{diagdiv}
  J_{\omega_{i,x}\omega_{i,x}} = (\alpha_{i} - 1 )^{2} \int d^{2} \omega
  \, e^{2\sigma} \: \frac{(\omega_{i,x} -\omega_{x})^{2}}{ |\omega -
    \omega_{i}|^{4}}
\end{equation}
and the behavior of $e^{2\sigma(\omega)}$ in the neighborhood of
$\omega_{i}$ is $e^{2\lambda_{i}} |\omega - \omega_{i}|^{2(\alpha_{i}
- 1)}$. However for $\alpha_{i} \rightarrow 1^{+}$ the term
(\ref{diagdiv}) does not diverge, actually goes to $0$ because of the
presence of the factor $(\alpha_{i} - 1 )^{2}$. Thus for these
diagonal matrix elements we must consider the analytic continuation
for $\alpha_{i} < 1$ and prove that such continuation is invariant
under $SL(2,C)$.

Setting $\delta_i=1-\alpha_i$, the analytic continuation of
\begin{equation}
 J_{\omega_{i,x}\omega_{i,x}} = \delta_i^{2} \int d^{2} \omega
  e^{2\sigma} \frac{(\omega_{i,x} -\omega_{x})^{2}}{ |\omega - \omega_{i}|^{4}}
\end{equation}
for $\delta_i >0$ is given by
\begin{eqnarray}
 & \delta_i^{2} \int d^{2} \omega
 \left[ e^{2\lambda_0} \prod_{j\neq
 i}|\omega-\omega_j|^{-2\delta_j}-e^{2\lambda_i} 
  e^{-|\omega-\omega_i|^2} \right]
 |\omega - \omega_{i}|^{-2\delta_{i}}
\frac{(\omega_{i,x} -\omega_{x})^{2}}{ |\omega -\omega_{i}|^{4}}
 &   \nonumber \\ & +   
{\pi\over 2} \Gamma(-\delta_i) \delta_i^2 e^{2\lambda_i}. &
\end{eqnarray}
We saw above that the transformation law
\begin{eqnarray}
\label{chain} 
J_{\omega_{i,a}\omega_{j,b}}  & = &
\sum_{cd}J'_{\omega_{i,c}\omega_{j,d}}
{\partial \omega'_{ic} \over \partial \omega_{ia}}
{\partial \omega'_{jd} \over \partial \omega_{jb}} +
\sum_{c}J'_{\omega_{i,c}\lambda}
{\partial \omega'_{ic} \over \partial \omega_{ia}}
{\partial \lambda' \over \partial \omega_{jb}}
  \\  & + &
\sum_{d}J'_{\lambda \omega_{i,d}}
{\partial \lambda' \over \partial \omega_{ia}} 
{\partial \omega'_{jd} \over \partial \omega_{jb}} +
J'_{\lambda \lambda}
{\partial \lambda' \over \partial \omega_{ia}} 
{\partial \lambda' \over \partial \omega_{jb}}
\nonumber
\end{eqnarray}
due to the invariance of the area $A$, holds in the convergence
region, i.e. for $i\neq j$ and if $i=j$ for $\delta_i <0$. As
${\partial \omega'_{ic} \over \partial \omega_{ia}}$ does not depend
on the $\alpha_j$ and ${\partial \lambda' \over \partial \omega_{ia}}$
is a linear function in the $\alpha_j$ (see eq.(\ref{misconfinv})),
the relation holds also for the matrix elements continued for
$\delta_i>0$. On the other hand the validity of eq.(\ref{chain}) implies
eq.(\ref{dettrans}). 

In $\det J$ one can separate the dependence on $\lambda_0$ and on the
harmonic ratios of the $\omega_i$ by writing
\begin{equation}
  \label{scomp}
  \sqrt{\det J} =  e^{3 N\lambda_{0} } W \prod_{i,j>i} |\omega_{i}
  - \omega_{j} |^{2\beta_{ij}}. 
\end{equation}
{}From eq.(\ref{dettrans}), under $SL(2,C)$ transformations
(\ref{misconfinv}), we have
\begin{equation}
  \sqrt{\det J'} = \sqrt{\det J} \, \prod_{i=1}^{N} |
  \omega_{i}c + d |^{4}
\end{equation}
which using eq.(\ref{scomp}) becomes
\begin{equation}
  \label{gentrans2}
  e^{3 N \lambda_{0}'} W' \,\prod_{i,j>i}
  | \omega_{i}' - \omega_{j}' |^{2 \beta_{ij}} =
  e^{3 N\lambda_{0}} W \,\prod_{i,j>i} |\omega_{i} - \omega_{j}|^{2
    \beta_{ij}} \prod_{i=1}^{N} | c \omega_{i} + d |^{4}.
\end{equation}
In order to have $W'=W$ the $\beta_{ij}$ must be chosen to satisfy
\begin{equation}
  \label{invreq}
  \displaystyle \prod_{i,j>i} | (\omega_{i} c + d
  ) (\omega_{j} c + d
  )|^{ -2\beta_{ij}} = \prod_{i} |\omega_{i}c + d |^{ 4 + 3 \delta_{i} N}
\end{equation}
i.e.
\begin{equation}
  \label{req}
  \sum_{j\neq i} \beta_{ij} = - ( 2 + \frac{3}{2} \delta_{i} N ).
\end{equation}
A particular solution of eq.(\ref{req}) is
\begin{equation}
  \label{betasol}
  \beta_{ij} = \frac{3}{2} \frac {N}{N-2} (\frac{2}{N-1}
  -\delta_{i} -\delta_{j}) - \frac{2}{N-1}.
\end{equation}
The conclusion is that $J$ can be written as
\begin{equation}
  \label{jform}
  J = e^{6N\lambda_{0}}\, W^2 \, \prod_{i, j>i} |\omega_{i} -
  \omega_{j}|^{4 \beta_{ij}}
\end{equation}
with $W$ a function of only $\omega_{i}$ and $\alpha_{i}$ which is
invariant under the full $SL(2,C)$ and thus function only of the
harmonic ratios of the $\omega_{i}$.

\section{Torus topology}
\label{torustopology}

The most general metric, modulo diffeomorphisms, is given by a flat
metric $\hat{g}_{\mu \nu}(\tau_{1}, \tau_{2})$ times a conformal
factor $e^{2\sigma}$.  $\tau_{1}$ and $\tau_{2}$ are the two
Teichm\"{u}ller parameters in terms of which, with $\tau=\tau_{1}+i
\tau_{2}$,
\begin{equation}
ds^{2}= dx^{2}+2\tau_{1}dx dy+ |\tau|^{2} dy^{2}
\end{equation}
and the fundamental region has been taken the square $ 0 \leq x <1, 0 \leq
y<1$. We recall the expression for the Green function of $\Box$ on the
torus \cite{minzub} using $\omega = x+ \tau y$
\begin{eqnarray}
& \Box G(\omega - \omega' |\tau) = \delta^{2}(\omega - \omega') -
\frac{1}{\tau_{2}} & \\ 
& G(\omega - \omega' | \tau)  = \frac{1}{2\pi} \log \left|
    \frac{\vartheta_{1} (\omega - 
\omega'| \tau )}{\eta(\tau)} \right| - \frac{(\omega_{y} -
\omega_{y}')^{2}}{2 \tau_{2}}  &
\end{eqnarray}
being $\vartheta_{1}(\omega | \tau)$ the Jacobi $\vartheta$--function
and
\begin{equation}
\displaystyle
\eta(\tau)=  e^{\frac{i\pi\tau}{12}} \prod_{n=1}^{\infty} [ 1 -
e^{2in\pi\tau} ] .
\end{equation}
{}From the written Green function and using
\begin{equation}
R(e^{2\sigma}\hat{g})=e^{-2\sigma}(R(\hat{g})-2 \hat{\Box}\sigma)
\end{equation}
it follows that the conformal factor $\sigma$, in presence of angular
deficits $2\pi(1-\alpha_{i})$ concentrated at the points $\omega_{i} $
is \begin{equation}
\label{sigmatoro}
\sigma (\omega) = \lambda_{0} + \sum_{i=1}^{N} (\alpha_{i} -1) \left\{
\log \left|  
    \frac{\vartheta_{1} (\omega - \omega_{i}| \tau )}{\eta(\tau)} \right|
  - \frac{\pi}{\tau_{2}}(\omega_{y} -\omega_{i,y})^{2}
\right\}.
\end{equation}

Thus the physical degrees of freedom are $3N$: in fact in addition to
the $2N$ $x_i,\, y_i$ we have $N-1$ independent angular deficits
($\sum_{i=1}^{N} (\alpha_{i} -1) = 0$), two Teichm\"{u}ller parameters
and $\lambda_{0}$, to which we must subtract the two conformal Killing
vectors of the torus. We have the same number of physical degrees of
freedom as the number of bones in a Regge triangulation of the torus
with $N$ vertices as it can be easily checked through the Euler
relation for a torus ($F + V = H = 3F/2$, from which $H=3V$).

The derivation of the Liouville action proceeds similarly as for the
sphere. Eqs.(\ref{beldelta}), (\ref{beldiff}), (\ref{belwrite}) are
unchanged. The main difference is given by the form of $\lambda_{i}$,
defined as before as $e^{2\lambda_{i}}= \lim_{\omega \rightarrow
\omega_{i}} (e^{2\sigma} | \omega - \omega_{i} |^{2(1 -
\alpha_{i})})$.  From eq.(\ref{sigmatoro}) we have for the torus
\begin{equation}
\lambda_{i} = \lambda_{0} + \sum_{j\neq i} 2\pi (\alpha_{j} -1)
G(\omega_{i} -\omega_{j} | \tau) + (\alpha_{i} -1) \log | 2\pi
\eta^{2}(\tau) |\, .  
\end{equation}
Now proceeding as after eq.(\ref{belwrite})
\begin{equation}
\sum_{i=1}^{N} \delta \alpha_{i} \, \lambda_{i} = \sum_{i=1}^{N}
(\alpha_{i} -1) \delta \left[ \sum_{j \neq i} 2\pi (\alpha_{j} -1)
G(\omega_{i} -\omega_{j} | \tau) \right] + \frac{1}{2} \delta \left[
\sum_{i=1}^{N} (\alpha_{i} -1)^{2} \log |2\pi \eta^{2} | \right]
\end{equation}
and then
\begin{equation} 
2 \sum_{i=1}^{N} \delta \alpha_{i} \, \lambda_{i} =
\delta \left[ \sum_{i=1}^{N} (\alpha_{i} -1)  \sum_{j \neq i} 2\pi
(\alpha_{j} -1 )
G(\omega_{i} -\omega_{j} | \tau) + \sum_{i=1}^{N} (\alpha_{i} -1)^{2}
\log |2\pi \eta^{2} | \right] \, ,
\end{equation}
having used $\sum_{i=1}^{N} (\alpha_{i} -1) = 0$ for the torus.

The final result is
\begin{equation}
\label{finale1}
\log
\sqrt{\frac{\det'(P^{\dag}P)}{\det(\phi_{a}, \phi_{b}) \det(\psi_{l},
\psi_{k})}} = 
\log \sqrt{\frac{\det'(P^{\dag}P)_{\hat{g}}}{\det(\phi_{a},
    \phi_{b})_{\hat{g}} \det(\psi_{k},\psi_{l})_{\hat{g}}}} +
\frac{26}{12} S_{l}
\end{equation}
with
\begin{eqnarray}
\label{torusliouv}
\lefteqn{
S_{l} =  \sum_{i,j\neq i} \frac{(1 - \alpha_{i})(1
    -\alpha_{j})}{\alpha_{i}} \left[  \log \left| \frac{\vartheta_{1}
    (\omega_{j} - \omega_{i}| \tau )}{\eta(\tau)} \right| 
- \frac{\pi}{\tau_{2}} (\omega_{i,y} - \omega_{j,y})^{2} \right]} & &
\nonumber \\ & & 
 + (\lambda_{0} - \log | 2\pi \eta^{2} | ) \sum_{i} (\alpha_{i} -
    \frac{1}{\alpha_{i}}) - \sum_{i} F(\alpha_{i}) \, .
\end{eqnarray}

This action is obviously invariant under the translation $\omega_{i}
\rightarrow \omega_{i} + a $ (with complex $a$) corresponding to the
two conformal Killing vectors of the tours, and compared to the sphere
topology is no longer invariant under dilatations, rotations and
special conformal transformations.

In the continuous limit eq.(\ref{torusliouv}) goes over to the well known
expression 
\begin{equation}  
\displaystyle
\label{contliouvtorus}
\frac{1}{8\pi} \int  d^{2}\omega \, d^{2}\omega' \;(\sqrt{g}
  R)_{\omega }
\frac{1}{\Box}(\omega, \omega')(\sqrt{g} R)_{\omega'}
\end{equation}

\subsection{Modular invariance}
\hfill

The partition function is given by eq.(\ref{contpart})
\begin{equation}
\label{toruspart}
\displaystyle
\int {\cal D}[\sigma] \frac{d^{2}\tau}{v(\tau)} 
\sqrt{\frac{\det'(P^{\dag}P)_{\hat{g}}}{\det(\phi_{a},
    \phi_{b})_{\hat{g}} \det(\psi_{k},\psi_{l})_{\hat{g}}}}
\det ( \psi_{m},\frac{\partial g}{\partial \tau_{n}} )
\; \displaystyle{e^{\frac{26}{12} S_{l}} }  
\end{equation}
where for a torus (for a detailed discussion see \cite{dauria})
$v(\tau) = \tau_{2} $ and
\begin{eqnarray}
\det (\phi_{a}, \phi_{b})_{\hat{g}} = \mbox{const } \tau_{2}^{2} \quad & \qquad
\det (\psi_{k}, \psi_{l}) = \mbox{const } \tau_{2}^{2} \\
\det (\psi_{m}, \frac{\partial g}{\partial \tau_{n}}) =
\mbox{const } \quad & \qquad [\det' 
(P^{\dag}P)]_{\hat{g}} =  \tau_{2}^{4} | \eta(\tau) |^{8}\, . \nonumber  
\end{eqnarray}
It is well known that the expression 
\begin{equation}
\frac{d^{2}\tau}{v(\tau)}
    \sqrt{\frac{\det'(P^{\dag}P)_{\hat{g}}}{\det(\phi_{a}, 
    \phi_{b})_{\hat{g}} \det(\psi_{k},\psi_{l})_{\hat{g}}}}
    \det ( \psi_{m},\frac{\partial g}{\partial \tau_{n}} ) =
    \mbox{const } \frac{d^{2}\tau}{\tau_{2}} |\eta(\tau)|^{4}
\end{equation}
is invariant under the modular transformation
\begin{equation}
\label{transmod}
\tau \longrightarrow \tau' = \frac{\tau a + b}{\tau c + d }
\end{equation}
with $(a,b,c,d) \in {\bf Z}$ and $ad -bc =1$.  Thus we are left to
prove the modular invariance of $\int {\cal D}[\sigma]
e^{\frac{26}{12} S_{l}}$.

This is achieved by accompanying the change in $\tau$ by a proper
change in the integration variable $\omega_{i}, \lambda_{0}$ given by
\begin{equation}
\label{modtrans}
\omega' = \frac{\omega}{\tau c + d} \qquad 
\qquad \qquad \lambda_{0}' = \lambda_{0} + \log |
  \tau c + d | \, .
\end{equation}
The last equation follows from the transformation of $\sigma(\omega;
\lambda_{0}, \omega_{i}, \alpha_{i}, \tau)$ under a change of
coordinates
\begin{eqnarray}
\lefteqn{
\sigma'(\omega'; \lambda_{0}, \omega_{i}, \alpha_{i}, \tau) \equiv 
\sigma(\omega(\omega'); \lambda_{0}, \omega_{i}, \alpha_{i}, \tau)
+ \log \left| \frac{d\omega}{d\omega'}\right| } & &  \\ & & = 
\sigma((\tau c + d) \omega'; \lambda_{0}, \omega_{i}, \alpha_{i}, \tau)
+ \log \left| \tau c + d \right| = 
\sigma(\omega'; \lambda_{0}', \omega_{i}', \alpha_{i}, \tau')
\nonumber
\end{eqnarray} 
keeping in mind  eq.(\ref{sigmatoro}) and the modular invariant
$G(\omega -\omega_{i} | \tau) = G(\omega' -\omega_{i}' | \tau')$. 

$S_{l}$, as given by eq.(\ref{torusliouv}), is invariant under transformations
(\ref{modtrans}), (\ref{transmod}) because of the just cited modular
invariance of the Green function and because
\begin{equation}
\eta \left (\frac{a\tau + b}{ c\tau + d} \right) =
e^{i\phi}(c\tau + d)^{\frac{1}{2}} \eta(\tau)
\end{equation}
compensates the change in $\lambda_{0}$.

Also we have 
\begin{equation}
\sqrt{J'} \prod_{i=1}^{N} d^{2}\omega_{i}' d\lambda_{0}'
\prod_{j=1}^{N-1} d\alpha_{j} =
\sqrt{J} \prod_{i=1}^{N} d^{2}\omega_{i} d\lambda_{0}
\prod_{j=1}^{N-1} d\alpha_{j}.
\end{equation}
In fact from the invariance of the area 
\begin{equation}
\label{areatorus}
\tilde{A} = e^{\lambda_{0} + \tilde{\lambda}_{0}} \int_{{\cal M}
  (\tau)} d^{2}\omega \prod_{j,i=1}^{N} e^{2\pi (\alpha_{i} -1)
  G(\omega - \omega_{i} | \tau) } e^{2\pi (\tilde{\alpha}_{i} -1)
  G(\omega - \tilde{\omega_{i}} | \tau) }
\end{equation}
under the transformations (\ref{transmod}), (\ref{modtrans}) it follows
\begin{equation}
  \prod_{i,j} \frac{\partial^{2} \tilde{A}}{\partial p_{i} \partial
  \tilde{p}_{j}} \, dp_{i} \, d\tilde{p}_{j} = \prod_{i,j}
  \frac{\partial^{2} \tilde{A}}{\partial p_{k}' \partial 
    \tilde{p}_{l}'} dp_{i}' \, d\tilde{p}_{j}' 
\end{equation}
This concludes our proof of the modular invariance of
eq.(\ref{toruspart}).

\section{Comparison with the smooth limit}
\label{comparison}

One might ask how far one can reach the results (\ref{primoris}),
(\ref{finale1}), starting from the Liouville action for a smooth
$\sigma$ 
\begin{equation}
\int d^2 \omega (-\sigma \Box \sigma +\mu^2 e^{2\sigma})=
\int d^2 \omega (-\sigma \Box \sigma) +\mu^2 A
\end{equation}
and then taking a proper singular limit.
One should construct a smooth surface depending on an invariant
parameter $\rho$ such that for $\rho\rightarrow 0$ it tends to our Regge
manifold with vertices at $\omega_i$ and angular openings
$\alpha_i$. This is a not trivial task; nevertheless we shall show
that a rough cut off procedure reproduces the main features of formulas
(\ref{primoris}), (\ref{finale1}).

For some aspect the problem is similar to that of electrostatics when
one takes the limit of a continuous distribution of charge to a point-like
distribution and the infinities arising from the self energies are
removed. The difference is that in our case it is difficult to
implement an exact cut off that is a geometric invariant in presence
of more than one singularity.

We shall regularize at an approximate level the tip of the cones with
segments of sphere (or pseudosphere) all with the same radius of
curvature $\rho/2$. The conformal factor describing a sphere of radius
$\rho/2$ around $\omega_{i}$ is
\begin{equation}
e^{2\sigma} = {k^2\over (1+({k\over \rho})^2 |\omega-\omega_i|^2)^2}
\end{equation}
for which $R=-2e^{-2\sigma} \Box \sigma ={8\over \rho^2}$.

$R$ is constant within a region $ |\omega-\omega_{i}| \leq r_{0}$,
where $r_{0}$ is related to the deficit angle by
\begin{equation}
r_{0}^2 = {\rho^2\over k^2} {1-\alpha\over 1+\alpha} \, .
\end{equation}
In presence of more than one singularity we shall impose that the
conformal factor $\sigma$ for  $ |\omega-\omega_i| > r_{0}$ goes over
to
\begin{equation}
\sigma = \lambda_0 +(\alpha_i-1) \log |\omega-\omega_i| +\sum_{j\neq
i}(\alpha_j-1) \log  |\omega_j-\omega_i| \, .
\end{equation}
Thus for $|\omega-\omega_{i}|=r_{0}$ we have
\begin{eqnarray}
\lefteqn{\log k =  
{1\over 2 \alpha_i}[ \log 4 + 2 \lambda_0 + (\alpha_i-1)(
\log\rho^2 +\log (1-\alpha_i))}  & & \\ & &-
(\alpha_i+1)\log(1+\alpha_i)  +2\sum_{j\neq i}(\alpha_j-1) 
\log |\omega-\omega_i|] \nonumber \, .
\end{eqnarray}

Integrating $\sigma \Box \sigma $ over the region $V$ around $\omega_i$ of
non vanishing curvature, we obtain
\begin{eqnarray}
\label{ultimas}
\lefteqn{ - \frac{1}{2\pi} \int_V \sigma \Box \sigma d^2 \omega } & & \\
& &
= - \sum_{i,j\neq i}
\frac{(1-\alpha_i)(1-\alpha_{i})}{\alpha_{i}} 
\log |\omega-\omega_{i}| + \sum_i f(\alpha_i) +
\lambda_0 ({1\over \alpha_i}-1) - \frac{1}{2\alpha_i}(1-\alpha_i)^2
\log \rho^{2} \, . \nonumber
\end{eqnarray}
Similarly one works with $\alpha >1$, i.e.\ with negative curvatures,
obtaining the same result.  The considered approximation, after
removing the divergent terms $\sim \log\rho^2$, misses with respect to
the exact expression (\ref{primoris}) the term $-\lambda_0 \alpha_i$
which is here replaced by $-\lambda_0$. This is obviously due the
approximate matching of the internal to the external metric. We stress
that, contrary to the exact expression (\ref{primoris}), the
approximate equation (\ref{ultimas}) is not invariant under $SL(2,C)$.

\section{Conclusions}

Applying to the Regge surfaces the conventional definition of
diffeomorphisms we have derived the analogous of the Liouville action
for the discretized case. Such results are exactly invariant at the
discretized level under the $SL(2,C)$ group for the sphere topology
and under translations and modular transformations for the torus
topology. In the continuum limit they go over to the usual continuum
results. For the sphere the action is given by eq.(\ref{primoris}) and
for the torus by eqs.(\ref{finale1}), (\ref{torusliouv}). The
integration measure for the conformal factor is provided for the
sphere topology by the determinant of the finite dimensional matrix
$J_{ij}$ eq.(\ref{jac2}) and for the torus topology by the analogous
expression obtained from the area $\tilde A$ given by
eq.(\ref{areatorus}). One could subject the partition function to
numerical computations. The action, even though non local (as one
expects from a functional determinant) is very simple especially for
the sphere topology. Probably the difficult part for a numerical
simulation is the evaluation of the finite dimensional determinant
$\det J_{ij}$.

The developed approach can be extended to any dimension \cite{pmpppp}.
However it appears very hard to provide an explicit expression for the
analogous of the functional determinant (\ref{siparte}); in fact for
$D\geq 3$ it is unlikely that the computation of such a determinant
can be reduced to the evaluation of local quantities as it happens for
$D=2$.

{\bf Acknowledgments}. We are grateful to A.\ D'Adda for pointing out
to us reference \cite{foer} and to M.\ Mintchev for interesting
discussions. 

\appendix

\section{Two dimensional continuum gravity}
\label{appc}

In the text we refer repeatedly to the continuum formulation; thus we
report here a concise derivation of the main formula following the
classical papers of \cite{allc}.  The starting point is the
formal functional integral
\begin{equation}
  Z = \int \frac{{\cal D}[g_{\mu\nu}]}{ V_{GC}} e^{-\int dx^2
    \sqrt{g}[\frac{T}{2}g^{ab}\partial_a X\partial_b X + \lambda R +
    \mu^2]}
\end{equation}
where $V_{GC}$ is the volume of the general coordinate transformation.
For the distance in the space of the metrics one adopts the De Witt
metric
\begin{eqnarray}
  \label{dewittm}
  & (\delta g_{\mu\nu}, \delta g_{\mu\nu})= \int \sqrt{g}~ d^2x~ \delta
  g_{\mu\nu} G^{\mu\nu,\mu'\nu'} \delta g_{\mu'\nu'} &
\\
&  G^{\mu\nu,\mu'\nu'}= g^{\mu\mu'} g^{\nu\nu'} + g^{\mu\nu'}
  g^{\nu\mu'} - C g^{\mu\nu} g^{\mu'\nu'} \, . &
  \nonumber
\end{eqnarray}
The most general metric can be written uniquely as
\begin{equation}
  g_{\mu\nu}= \tilde f^*(e^{2\sigma}\hat g(\tau))_{\mu\nu}=
  g_{\mu\nu}(\sigma,\tau,\tilde{f})
\end{equation}
where $\tilde f$ is the general diffeomorphism orthogonal to the action
of a conformal transformation. A variations $\delta g_{\mu\nu}$ of the
metric can be decomposed into a Weyl transformation, plus a coordinate
transformation plus a change in the Teichm\"uller parameters
\begin{eqnarray}
  \lefteqn{
  \delta g_{\mu\nu} =\nabla_\mu\tilde \xi_\nu + \nabla_\mu\tilde
  \xi_\nu + 2 \tilde{f}^*\delta\sigma ~g_{\mu\nu} + \frac{\partial 
  g_{\mu\nu}}{\partial \tau_i} \delta \tau_i } & &  \\ & & =
  P[\tilde \xi+ \frac{1}{ P^\dagger P}P^\dagger k^i_{\mu\nu}
  \delta\tau_i] +
  g_{\mu\nu} ( 2 f^*\delta\sigma + \nabla_g \tilde \xi +
  \frac{g^{\alpha\beta}}{ 2} \frac{\partial g_{\alpha\beta}}{ \partial
    \tau_i} \delta \tau_i)+ (1 - P \frac{1}{P^\dagger P}P^\dagger)
  k^i_{\mu\nu} \delta \tau_{i} \nonumber
\end{eqnarray}
where
\begin{equation}
  k^i_{\mu\nu} = \frac{\partial g_{\mu\nu}}{\partial \tau_i} -
  \frac{g_{\mu\nu}}{2} g^{\alpha\beta} \frac{\partial 
  g_{\alpha\beta}}{\partial \tau_i} \, .
\end{equation}
We shall find
\begin{equation}
  {\cal D}[g] = {\cal D}[\tilde f] \, {\cal D}[\sigma] \, d\tau_i \,
  J(\sigma,\tau)
\end{equation}
where $J$ does not depend on $\tilde f$. If now one integrates a
diff-invariant quantity $\cal F$, we have
\begin{equation}
  \label{varchange}
  \int \! {\cal D}[g] \: {\cal F} =\int \! {\cal D}[f] \int \! \frac{{\cal
    D}[\tilde 
    f]}{{\cal D}[f]} \: {\cal D}[\sigma] \: d\tau_i \: J(\sigma,\tau)
    \; {\cal F} \, .
\end{equation}

In order to find $J$ let us write, using the standard normalization
\begin{eqnarray}
  \lefteqn{ 1 = \int {\cal D} [\delta g_{\mu\nu}] e^{-\frac{1}{2}(\delta
    g_{\mu\nu}, \delta g_{\mu\nu})}  = \int {\cal D} [\tilde{\xi}_\mu]
  {\cal D} [\delta \sigma] d\delta\tau_i J(\sigma,\tau) } & & \\ 
  & & \times
  \exp \left[ -\frac{1}{2} (\delta\sigma,\delta\sigma)-\frac{1}{2}(P\tilde
    \xi,P\tilde \xi) - \frac{1}{2} ((1 - P \frac{1}{P^\dagger
      P}P^{\dagger})k^i_{\mu\nu},((1 - P \frac{1}{P^\dagger
    P}P^{\dagger}))k^i_{\mu\nu} 
    )\delta \tau_i \delta \tau_j \right] \nonumber
\end{eqnarray}
where in the first two terms, invariance of the measure on the tangent
space under translations has been used and $J$ does not depend on $f$,
due to diff-invariance; $1- P \frac{1}{P^\dagger P}P^{\dagger}$ is the
projector on the zero modes $\psi_l$ of $P^\dagger$. Taking into
account that $\psi_l$ are traceless, we have
\begin{equation}
  \label{normg}
  \displaystyle
  1 = J(\sigma, \tau) [\mbox{det}'(P^\dagger P)]^{- \frac{1}{2}}
  [\det(\psi_m,{\partial g\over\partial\tau_n})]^{-1}
  [\det(\psi_k,\psi_l)]^{\frac{1}{2}} \, . 
\end{equation}
Finally one must compute
\begin{equation}
  \int {\cal D}[\tilde f]/{\cal D}[f] \, .
\end{equation}
This can be achieved by the following change of variable
\begin{equation}
  {\cal D}[f]= {\cal D}[\tilde f]\Pi dw_c \; K
\end{equation}
where $w_c$ are the normal coordinates associated to the $N$ conformal
Killing vectors $\phi_a$ (which satisfy $P(e^{2\sigma}\phi_a)=0$ and
as such do not depend on $\sigma$, see eq.(\ref{opland})). In order to
find $K$ one computes on the space tangent to the diffeomorphisms
\begin{eqnarray}
  \label{normx}
\lefteqn{ 1 = \int {\cal D} [\xi] e^{ - {1\over 2}(\xi,\xi)_g}} & & \\
 & & = \int \! {\cal D}[\tilde \xi] \, \prod_c d\delta w_{c} \, K \,
 \exp \left[ - {1\over 2}(\tilde \xi, \tilde \xi)_{g}- {1\over
 2}(\phi_a,\phi_b)_{g} \delta w_a \delta w_b \right] = \mbox{const }
 K[\det (\phi_a,\phi_b)]^{-{1\over 2}} \, .
 \nonumber
\end{eqnarray}
Thus
\begin{equation}
  \int {\cal D}[f] = \int dw_c \int {\cal D}[\tilde f] \:
  \det(\phi_a,\phi_b)^{1\over 2}= v(\tau) \int {\cal D}[\tilde f] \:
  [\det(\phi_a,\phi_b)]^{1\over 2} \, ,
\end{equation}
being $v(\tau)$ the volume of the group generated by the conformal
Killing vectors.  
{}From eqs.(\ref{varchange}), (\ref{normg}), (\ref{normx}) one can write
\cite{allc}
\begin{equation}
  \label{contpart}
  \int {\cal D}[g] {\cal F} = \int {\cal D}[f] \: \int {\cal D}[\sigma] 
  {d\tau_i\over v(\tau)} 
  \left[ {\det'(P^\dagger P)\over
    \det(\phi_a,\phi_b)\det(\psi_k,\psi_l)} \right]^{1\over 2} \det
  (\psi_m,{\partial g\over \partial\tau_n}) \, .
\end{equation}
We recall that $v(\tau)$ and $\det(\psi_m,{\partial g\over
  \partial\tau_n})$ do not depend on $\sigma$ but only on the
Teichm\"uller parameters $\tau_i$, while the remaining square root is
the exponential of the Liouville action multiplied by the same
quantity at $\sigma =0$.

\section{Asymptotic expansion of the heat kernel}
\label{appa}

In this appendix we summarize the computation of $\Xi_{K}(0)$, where
$\Xi_{K}(s)$ is given by
\begin{equation}
  \Xi_{K}(s) = {1\over \Gamma(s)} \int_0^\infty t^{s-1} \mbox{Tr}(
  e^{-tK})dt,
\label{zetamode}
\end{equation}
employing the direct method of Cheeger \cite{chee} that does not involve
contour integrals.

The trace Tr involves also the summation on the angular momentum
$m$. In \cite{chee} such a sum is split into two parts, which give
rise to contributions to $\Xi_{K}(s)$ that are analytic in two non
overlapping vertical strips in the complex $s$-plane.  $\Xi_{K}(s)$ is
defined as the sum of the two analytic continuations. $\Xi_{K}(0)$ is
given by the constant term in the asymptotic expansion of the trace of
the heat kernel. Furthermore in \cite{chee} it is proven that
\begin{equation}
  \Xi_{K}(s) = {\rm Tr} (K^{-s})
\end{equation}
as can be expected from eq.(\ref{zetamode}).

In our case with $K=L^\dagger L$, it is easy to prove from the choice
of the eigenfunctions given in the text, that
\begin{equation}
  \mbox{Tr} (K^{-s}) = \int_{0}^{R_{0}} R \,dR \sum_m \int_{0}^{\infty}
  J^2_{\nu_m}(\lambda R) \lambda ^{1-2s} d\lambda
\end{equation}
where for clearness sake we have explicitly indicated a space cut-off
$R_0$, that will disappear in the value of $\Xi_{K}(0)$.

Using the relation
\begin{equation}
  \int_{0}^{\infty} J^2_{\nu}(\lambda R) \lambda^{1-2s} d\lambda =
  R^{2(s-1)} \frac{\Gamma(\nu-s+1)\Gamma(s-1/2)}{2 \sqrt{\pi}
  \Gamma(\nu+s)\Gamma(s)} 
\end{equation}

we obtain

\begin{equation}
  \Xi_{K}(s)= {R_0^{2s}\over 2s \Gamma(s)} {\Gamma(s-1/2)\over
    2\sqrt{\pi}}\sum_m {\Gamma(\nu_m-s+1)\over \Gamma(\nu_m+s)}.
\end{equation}

For $s\rightarrow 0$ we have the asymptotic
expansion \cite{chee}
\begin{equation}
  \frac{\Gamma(\nu-s+1)}{\Gamma(\nu+s)}=\nu^{1-2s}(
  1+s\sum_{j=1}^{\infty} \frac{B_{2j}}{j}\nu^{-2j}+O(s^2))
\end{equation}
with $B_j$ the Bernoulli numbers (for the Bernoulli numbers and
polynomials we use the notation of \cite{erdely}). In the limit
$s\rightarrow 0$, $\nu_m={m+\delta\over \alpha}$ for $m\ge 0$ and
$\nu_m= -\frac{m+\delta}{\alpha}$ for $m < 0$, taking into account
that $\zeta(u,a)$ has a simple pole only at $u=0$ with residue 1, we
have 
\begin{eqnarray}
  \label{sommappa}
  \lefteqn{
  \Xi_{K}(s)\rightarrow -{1\over 2} \left\{ \sum_{n=0}^\infty [
  ({n+\delta\over\alpha})^{1-2s}+ s ({n+\delta\over\alpha})^{-1-2s}B_{2}]
  \right. } & &  \\ & &
\left. +  \sum_{n=1}^\infty [ ({n-\delta\over\alpha})^{1-2s}+ s
  ({n-\delta\over\alpha})^{-1-2s}B_{2}] \right\} \nonumber \\ & &
=  -{1\over 2}[ \alpha^{2s-1}\zeta(2s-1,\delta) + B_{2} s\,
  \alpha^{2s+1}\zeta(2s+1,\delta) + \alpha^{2s-1}\zeta(2s-1,1- \delta)
   \nonumber \\ & & 
+ B_{2} s\, \alpha^{2s+1}\zeta(2s+1,1- \delta)].
\nonumber
\end{eqnarray}
Recalling that $B_{2} = {1\over 6}$, $\lim_{s\rightarrow 0}
s\zeta(1+2s,\delta)={1\over 2}$ , $\zeta(-1,v)= -{B_2(v)\over 2} =
-{1\over 2}(v^2-v+{1\over 6})$ we have \cite{dowker}
\begin{equation}
  \Xi_{K}(0)={1-\alpha^2\over 12 \alpha} + {\delta(\delta-1)\over 2
    \alpha}.
\label{c0k}
\end{equation}

If instead of $K=L^{\dag}L$ we consider $H=LL^{\dag}$ with
eigenfunctions given by $h_{\lambda}= L\xi_{\lambda}$, due to the
tensor character of $h_{\lambda}$ and the corresponding measure
(\ref{tenmis}) we have
\begin{equation}
\Xi_{H} (s) = \mbox{Tr}(H^{-s}) = \int_{0}^{R_{0}} R \, dR
\int_{0}^{\infty} J^{2}_{\gamma_{m}} (\lambda R) \lambda^{1 - 2s} d\lambda
\end{equation}
with $\gamma_{m} = \nu_{m} +1$ for $\nu_{m} = \frac{m + \alpha
  -1}{\alpha}$ and $\gamma_{m} = \nu_{m} -1$ for $\nu_{m} = - \frac{ m
  + \alpha -1}{\alpha}$. It corresponds to change $\delta$ into
  $\delta + \alpha$ in eq.(\ref{c0k}), thus obtaining
\begin{equation}
  \Xi_{H}(0)={1-\alpha^2\over 12 \alpha} + \frac{(\delta + \alpha)(\delta +
    \alpha -1)}{2 \alpha}.
\label{c0h}
\end{equation}
We remark that $\Xi_{K}(0)$ and $\Xi_{H}(0)$ are the constant
coefficients $c_{0}^{K}$ and $c_{0}^{H}$ in the asymptotic expansion
of the trace of the heat kernels ${\cal K}$ and ${\cal H}$.

\section{Regularization of conical singularities}
\label{appsphere}

In the text we gave the general analysis of the self-adjoint
extensions of the $L^{\dag} L$ and we showed that the imposition of
the validity of the Riemann-Roch relation fixes a well defined form on
the eigenfunctions for this operator in the interval $\frac{1}{2} <
\alpha<2$.  Here we shall derive briefly the same result using a more
conventional method, i.e. that of regularizing the conical
singularity.  The regulator we shall use is to replace the tip of the
cone with a segment of a sphere (or of the Poincar\'e pseudo-sphere)
connected smoothly to the remaining part of the cone.  In such a scheme
the eigenfunction problem can be solved exactly.  The form of the
eigensolutions on the cone will be fixed by taking the limit where the
radius of the sphere (or pseudo--sphere) tends to zero, keeping the
integrated curvature fixed.

A sphere of radius ${1\over 2}\rho$ of constant curvature $R=-2
e^{-2\sigma} \Box \sigma = 8\rho^{-2}$ or the pseudo-sphere of
constant curvature $R=-8\rho^{-2}$ are described on the
$\omega$--plane by the conformal factor $e^{2\sigma}=(1\pm u \bar
u)^{-2}$ with $u=\omega/\rho$. Similarly a cone is described by the
conformal factor $e^{2\sigma} = c^2 (\omega \bar{\omega})^{\alpha -
1}$. The radius at which the sphere connects to the cone will be
denoted by $r_0 = \rho v_{0}$, ($r = |\omega|$ and $v=|u|$).
$c$ is fixed by the matching condition
\begin{equation}
(1\pm \left(\frac{r_{0}}{\rho}\right)^{2})^{-2} = c^2
(r_{0}^{2})^{\alpha - 1},
\end{equation}
from which we obtain $c = \rho^{1-\alpha} \displaystyle{
\frac{v_{0}^{1-\alpha}}{1 \pm v_{0}^{2}} }$.
The integrated curvature on the segment of the sphere (pseudo--sphere)
between $r=0$ and $r=r_{0}$ is given by $\pm 8\pi v_{0}^{2}/(1\pm
v_{0}^{2})$. Imposing this quantity to be equal to $4\pi(1 - \alpha)$
(i.e. to the curvature concentrated on the tip of the cone) fixes the
value of $r_{0} = \rho v_{0}$ to
\begin{equation}
v_{0}^{2} ={1-\alpha \over 1+\alpha} \qquad \quad \mbox{for the 
sphere } (0<\alpha<1)  \label{v0sfera}
\end{equation}
and
\begin{equation}
v_{0}^{2} ={\alpha-1  \over \alpha+1} \qquad \quad \mbox{for the
pseudo-sphere } (1<\alpha).   \label{v0pseudo}
\end{equation}
{}From eq.(\ref{v0sfera}) and eq.(\ref{v0pseudo}) we see that a segment of
sphere or pseudo--sphere with given integrated curvature is described
by a fixed value $v_{0}$, corresponding to $r_{0} = \rho v_{0}$.

We shall now find the eigensolutions of $L^{\dag}L$ on the sphere or
pseudo--sphere and then we shall connect them smoothly to those on
the cone.  Let us consider first the case of positive curvature.
Using eqs.(\ref{opland}), (\ref{opl}), (\ref{oplc}) the general eigenvalue
equation
\begin{equation}
e^{-2\sigma} L^{\dag} e^{2\sigma} L e^{-2\sigma} \: \xi = \lambda^{2}
\xi 
\end{equation}
becomes
\begin{equation}
\left\{ x ( 1 + x)^{2} \frac{d^{2}}{dx^{2}} + (1 + x) (3x +1)
  \frac{d}{dx} - \frac{m^{2} (1+x)^{2}}{4 x} + m (1 + x) + 2 +
  \lambda^{2} \rho^{2} \right\} \xi^{(m)} = 0 
\end{equation}
where we have set $\xi = e^{im\phi} \xi^{(m)}$ and $x=r^{2} =
\bar{\omega} \omega$. Such an equation has three regular singular
point and thus can be solved by standard methods \cite{erdely}. 
We find
\begin{eqnarray}
m\geq0 \qquad \qquad & \xi^{(m)} =
\displaystyle{\frac{v^{m}}{(1+v^{2})^{2}}}  \quad_{2}F_{1} 
(\gamma_{1} + 2, 1 - \gamma_{1}; 1 + m ; \frac{v^{2}}{1+ v^{2}}) \\
m < 0 &  \xi^{(m)} = v^{-m} \quad_{2}F_{1} (\gamma_{1}, -1
- \gamma_{1}; 1 - m ; \displaystyle{\frac{v^{2}}{1+ v^{2}}})
\nonumber
\end{eqnarray}
where $\gamma_{1} = \frac{1}{2}(-1+\sqrt{9+4(\rho\lambda)^{2}}~)$.
For $\rho^2 = 0$ they reduce to
\begin{eqnarray}
m \geq 0 &   \displaystyle{\frac{v^{m}}{ (1 +
    v^{2})^{2}}} \\ 
m < 0 \quad \qquad &  \displaystyle{\frac{v^{-m}}{ (1 +
v^{2})^{2}}} 
[ (1 + v^{2} )^{2} - {2 \over 1-m} v^{2} ( 1 +
v^{2}) + {2\over (1-m)(2-m)} v^{4} ].
\nonumber
\end{eqnarray}
On the pseudo-sphere we have
\begin{eqnarray}
m\geq0 \qquad \qquad \quad & \xi^{(m)} = \displaystyle{ \frac{v^{m}}{(1 -
    v^{2})^{2}}} ~_{2}F_{1}  
(\gamma_{2} + 2, 1 - \gamma_{2}; 1 + m ; \frac{v^{2}}{v^{2} - 1} ) \\
m < 0 & \xi^{(m)} = v^{-m} \quad_{2}F_{1} (\gamma_{2}, -1
-\gamma_{2}; 1 - m; \displaystyle{\frac{v^{2}}{v^{2} - 1 }})
\nonumber
\end{eqnarray}
where $\gamma_{2} = \frac{1}{2}(-1+\sqrt{9-4(\rho\lambda)^{2}}~)$.
For $\rho^2 = 0$ we obtain
\begin{eqnarray}
m \geq 0 &   \displaystyle{\frac{v^{m}}{ (1 - v^{2})^{2}}} \\ 
m < 0 \qquad \quad &  \displaystyle{\frac{v^{-m}}{ (1 -
    v^{2})^{2}}}  [(1 - v^{2} )^{2} + {2\over 1-m}
v^{2} ( 1 - v^{2}) + {2 \over (1-m)(2-m)} v^{4} ].
\nonumber
\end{eqnarray}

As we know from eq.(\ref{conoeq}) the general eingensolution on the cone for
orbital angular momentum $m$ has the form
\begin{equation}
\xi_{\rm ext}^{(m)} = 
  v^{\alpha -1} \left[ a(\rho) J_{\gamma} 
(2\rho\lambda p v^{\alpha})  + b(\rho) J_{-\gamma} (2\rho\lambda p v^{\alpha}) \right]
\end{equation}
where $\gamma=\frac{m+\alpha -1}{\alpha}$ and
$p =\frac{v_{0}^{1-\alpha}}{\alpha(1 \pm v_{0}^{2})}$.
The coefficients $a(\rho)$ and $b(\rho)$ are fixed by
requiring the continuity of the logarithmic derivative of
$e^{-2\sigma} \xi$ with respect to
$\bar{\omega}$ at $\mid \! \omega\!\! \mid = r_{0}$.
In fact from the structure of eigenvalue equation $e^{-2\sigma}
\frac{\partial}{\partial\omega} e^{2\sigma}
\frac{\partial}{\partial\bar{\omega}} e^{- 2\sigma} \xi = - \lambda^{2}
\xi$, we see that failing to satisfy such a condition would produce a
singular contribution at the matching point. 

Let us see what is the form of the eigenfunctions on the cone fixed
by the matching condition in the limit when the regulator $\rho$ tends to 0.

We consider first $m\geq0$. For small $\rho$ the interior
solution multiplied by the factor $e^{-2\sigma}$ becomes
\begin{equation}
e^{-2\sigma} \xi_{int} = u^{m} \left[ 1 + (\rho\lambda)^{2} f \left(
\frac{u\bar{u}}{1 \pm u\bar{u}}  \right) +
O((\rho\lambda)^{4}) \right]
\end{equation}
while the exterior solution multiplied by the conformal factor
$e^{-2\sigma}$ becomes
\begin{equation}
  (\rho\lambda)^{\gamma} a(\rho) u^{m} \left[ c_{0} + c_{1} (\rho\lambda)^{2}
  (u\bar{u})^{\alpha} +O((\rho\lambda)^{4}) \right] +
  (\rho\lambda)^{-\gamma} b(\rho) \bar{u}^{-m} \left[  (u
  \bar{u})^{1-\alpha}  + O((\rho\lambda)^2) \right]  \label{int1}.
\end{equation}

We notice that the lowest order in $\rho\lambda$ in the first term of
eq.(\ref{int1}) has vanishing derivative with respect to
$\bar{\omega}$. Thus the continuity of the logarithmic derivative for
small $\rho\lambda$ takes the form
\begin{equation}
  \frac{1}{\rho} k(\rho\lambda)^{2} = \left. \frac{1}{\rho} \frac{a(\rho)
    c_{1} (\rho\lambda)^{2} +
    b(\rho) c_{2} (\rho\lambda)^{-2\gamma }}{ a(\rho) c_{3} + c_{4}
      b(\rho) (\rho\lambda)^{-2\gamma} } \right.
\end{equation}
which gives
\begin{equation}
  \left. \frac{b(\rho)}{a(\rho)} =  \frac{ (\rho\lambda)^{2 + 2 \gamma} (c_{1}
    - k c_{3}) }{k (\rho\lambda)^{2} c_{4} - c_{2}}  \right. .
\label{b1}
\end{equation}
Thus for $m\geq 0$ we see that for $2 + 2 \gamma > 0$, i.e.\ $\alpha >
\frac{1}{2}$, $b(\rho)$ vanishes when the regulator is removed at
constant integrated curvature.  

Similarly one can deal with $m<0$. In this case the derivative of
the interior solution multiplied by the conformal factor tends  to a
finite limit for $(\rho\lambda)^{2} \rightarrow 0$ and the analog of
equation (\ref{b1}) is
\begin{equation}
\left. \frac{a(\rho)}{b(\rho)}  =  \frac{  (c_{5} - k_{1} c_{7})
  (\rho\lambda)^{-2\gamma} }{k_{1} c_{8} - c_{6}(\rho\lambda)^{2}}
  \label{b2} \right. .
\end{equation}
Thus for $m<0$ we have $a(\rho) \rightarrow 0$ for $\gamma<0$, i.e.\
for $\alpha<2.$

Thus for the opening of the cone $\alpha$ with $\frac{1}{2} <
\alpha < 2$, as the regulator is removed, only the term
$J_{\frac{m+\alpha -1}{\alpha}}$ survives for $m\geq 0$, while for
$m<0$ the surviving term is $J_{-\frac{m+\alpha -1}{\alpha}}$. This is
exactly the same result obtained in sect.\ \ref{sezadj} by imposing
the Riemann--Roch relation.

\section{Integral representation of the heat kernel}
\label{appintrep}
 
We give here the expressions of the integrals appearing in
eq.(\ref{daprima}). Let us consider \cite{dowker,pmppp}:
\begin{equation}
\label{intrephk}
K_{\alpha,\delta} ({\bf x}, {\bf x}';t) = \frac{1}{4\pi t}
  e^{-\frac{|{\bf x} - {\bf x}'|^{2}}{4t}} + \frac{1}{16 i \pi^2\alpha
  t} \int_{\Gamma} \!  d\zeta \; e^{-\frac{1}{4t} (x^2 + {x'}^2 - 2 x x' \cos
  \zeta)} \; \; \frac{e^{\frac{i}{2\alpha}(\zeta + \phi -
  \phi')(2\delta - 1)}}{\sin \frac{\zeta + \phi - \phi' }{2\alpha}}
  \label{green1}
\end{equation}
where $\phi$ is the polar angle associated to the vector ${\bf x}$ and
the where the integration contour $\Gamma$ is composed of the two
lines which go from $-\pi - i\infty$ to $-\pi + i\infty$ and from
$\pi+i\infty$ to $\pi-i\infty$.  We have ${\cal
K}_\alpha({\bf x}, {\bf x}';t) = K_{\alpha,\alpha-1}
({\bf x},{\bf x}';t)$ 
and ${\cal H}_\alpha({\bf x}, {\bf x}'; t) = K_{\alpha,
2\alpha-1}({\bf x}, {\bf x}';t)$.
Evaluating the kernel (\ref{intrephk}) at ${\bf x}' = {\bf x}$ we obtain
\begin{equation}
K_{\alpha,\delta}({\bf x}, {\bf x};t) = {1\over 4\pi t} + \frac{1}{16 i
  \pi^2\alpha t} \int_{\Gamma} \! 
d\zeta \;
e^{-\frac{1}{2t} r^2(1- \cos \zeta)} \; \;
 \frac{e^{\frac{i}{2\alpha}\zeta(2\delta - 1)}}{\sin \frac{\zeta}{2\alpha}} 
\end{equation}
from which
\begin{eqnarray}
\label{ultimafor}
\lefteqn{
\mbox{Finite}_{\epsilon \rightarrow 0} 
\int \! d^{2}x \, \log(\alpha | {\bf x}|)
K_{\alpha,\delta}({\bf x}, {\bf x};\epsilon) } & & \\ & & =
(\log \alpha -\frac{\gamma_{E}}{2}) \left[ \frac{\delta(\delta -1)}{
2\alpha} + \frac{1-\alpha^{2}}{12\alpha} \right] - 
\frac{1}{16 i \pi}\int_{\Gamma} \!
d\zeta \;
 \frac{e^{\frac{i}{2\alpha}\zeta(2\delta - 1)}}{\sin \frac{\zeta}{2\alpha}} 
\frac{\log(1-\cos\zeta)}{1 -\cos \zeta}. \nonumber
\end{eqnarray}
It is easily checked that the last integral converges for $|2\delta -1|
<2\alpha +1$. We notice that in our range ${1\over 2}<\alpha<2$, the
above inequality is satisfied both for $\delta = \alpha-1$ and for
$\delta = 2\alpha-1$. One can use the method of \cite{aur} to write
the integral of expression (\ref{ultimafor}) in
$\frac{\delta\alpha}{\alpha}$ as a single integral of real  function.

\bibliographystyle{plain}

\end{document}